\documentclass[numbers,preprint]{sigplanconf}
\usepackage{array,etoolbox}
\usepackage{amssymb}
\usepackage{amsmath}
\usepackage{url}
\usepackage{adjustbox}
\usepackage{flushend}
\usepackage{xspace}
\usepackage{mathtools}

\usepackage{stmaryrd}
\newsavebox\eglstbox
\newsavebox\egguardbox
\newsavebox\egestructbox
\newsavebox\egproductbox

\usepackage{listings}
\usepackage{subcaption}
\usepackage{float}
\usepackage{tikz, pgfplots}
\usetikzlibrary{decorations.pathreplacing}
\usetikzlibrary{calc}
\usetikzlibrary{arrows}			
\usetikzlibrary{automata}		
\usetikzlibrary{shapes,snakes, fit}

\usetikzlibrary{external}

\usetikzlibrary{shapes.multipart, chains, positioning, shapes.geometric}
\usetikzlibrary{shapes.symbols}
\usetikzlibrary{decorations.markings}
\usetikzlibrary{intersections}

\tikzset{
	box/.style=	{rectangle, draw=blue, thick, minimum width = 1.5em,
                	minimum height = 1.7ex, inner sep = 0
				},
	circ/.style={circle, draw=gray, minimum size = 3pt, 
					inner sep = 0
				},
	tsobuff/.style=	{rectangle, draw=blue, minimum width = 1em,
						minimum height = 0.75ex, inner sep = 0
					},
	tsobuffhead/.style=	{rectangle, draw=blue, minimum width = 1em,
                			minimum height = 1.75ex, inner sep = 0
						},
	psobuff/.style=	{rectangle split,  inner sep = 0,
						rectangle split horizontal,rectangle split parts=3, 
						draw=blue,
						text height=0.5ex, text width=0.5em,
						minimum width= 0.5em, minimum height = 0.75ex, 
					},
	psobuffhead/.style=	{rectangle split,  inner sep = 0,
							rectangle split horizontal,
  							text height=0.5ex, text width=0.5em,
							rectangle split parts=3, draw=blue,
							minimum width= 0.5em, minimum height = 1.75ex, 
						}
}

\usepackage{macros}

\usepackage{hyperref}		
\usepackage[capitalise]{cleveref}

\usepackage[firstpage]{draftwatermark}
\SetWatermarkText{\hspace*{-0.01em}\raisebox{4.45ex}{\includegraphics[scale=1]{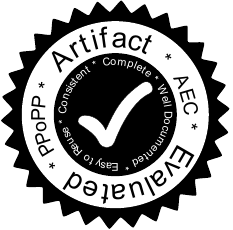}}}
\SetWatermarkAngle{0}

\begin{document}

\ifx 0
\conferenceinfo{PPoPP '16}{March 12-16, 2016, Barcelona, Spain}
\copyrightyear{2016}
\copyrightdata{978-1-4503-4092-2/16/03}
\reprintprice{\$15.00}
\copyrightdoi{2851141.2851165}
\publicationrights{transferred}

\CopyrightYear{2016}
\conferenceinfo{PPoPP '16,}{March 12-16, 2016, Barcelona, Spain}
\isbn{978-1-4503-4092-2/16/03}\acmPrice{\$15.00}
\doi{http://dx.doi.org/10.1145/2851141.2851165}
\fi

\title{The Virtues of Conflict: Analysing Modern Concurrency}

\authorinfo{Ganesh Narayanaswamy}
           {University of Oxford}
           {ganesh.narayanaswamy@cs.ox.ac.uk}

\authorinfo{Saurabh Joshi}
	    {IIT Guwahati}
	    {sbjoshi@iitg.ernet.in}

\authorinfo{Daniel Kroening}
           {University of Oxford}
           {kroening@cs.ox.ac.uk}
\ifx 0
\authorinfo{}
           {}
           {}
\fi
\maketitle

\begin{abstract}
Modern shared memory multiprocessors permit reordering of memory operations
for performance reasons.  These reorderings are often a source of subtle
bugs in programs written for such architectures.  Traditional approaches to
verify weak memory programs often rely on interleaving semantics, which is
prone to state space explosion, and thus severely limits the scalability of
the analysis.  In recent times, there has been a renewed interest in
modelling dynamic executions of weak memory programs using partial orders. 
However, such an approach typically requires ad-hoc mechanisms to correctly
capture the data and control-flow choices/conflicts present in real-world
programs.  In this work, we propose a novel, conflict-aware, composable,
truly concurrent semantics for programs written using C/C++ for modern weak
memory architectures.  We exploit our symbolic semantics based on general
event structures to build an efficient decision procedure that detects
assertion violations in bounded multi-threaded programs.  Using a large,
representative set of benchmarks, we show that our conflict-aware semantics
outperforms the state-of-the-art partial-order based approaches.

\end{abstract}

\category{}{F1.2}
{Modes of Computa\-tion}
[Parallelism and concurrency]

\terms
Verification


\section{Introduction}
\label{sec:intro}
 
\subsection{Problem Description}

Modern multiprocessors employ a variety of caches, queues and buffers to
improve performance.  As a result, it is not uncommon for write operations from a
thread to be not immediately visible to other threads in the system.  Thus,
writes from a thread, as seen by an external observer, may appear to have
been reordered.  The specifics of these processor-dependent 
reorderings are presented to programmers as a contract, called the
\defw{memory model}.  A memory model dictates the order in which operations
in a thread become visible to other threads~\cite{wmm:adve}.  Thus, given a
memory model, a programmer can determine which values could be returned by a
given read operation.

While most developers are aware that instructions from two different threads
could be interleaved arbitrarily, it is not atypical for a programmer to
expect statements \emph{within} one thread to be executed in the order in
which they appear in the program text, the so called \defw{program
order}~(\progo).  A memory model that guarantees that instructions from a
thread are always executed in program order is said to offer sequential
consistency~(\seqc)~\cite{lamportsc}.  However, none of the performant,
modern multiprocessors offer \seqc: instead, they typically implement what
are known as \defw{relaxed} or \defw{weak memory models} (R/WMM), by
relaxing/weakening the program order for performance reasons.  In general,
the weaker the model, the better the opportunities for performance
optimisations: the memory model alone could account for 10--40\% of the
processor performance in modern CPUs~\cite{mmperf}.

Such weakenings, however, not only increase performance, but also lead to
intricate weak-memory artefacts that make writing correct multiprocessor
programs non-intuitive and challenging.  A~key issue that compounds and
exacerbates this difficulty is the fact that weak-memory bugs are usually
non-deterministic: that is, weak memory defects manifest \emph{only} under
very specific, often rare, scenarios caused by a particular set of write
orderings and buffer configurations.  Although all modern architectures
provide {\em memory barriers} or {\em fences} to prevent such
relaxation from taking place around these barriers, the placement of fences
remains a research topic~\cite{jk2015-fm, wmm:fence:ver, wmm:fence:sc1,
wmm:fence:hw5, wmm:fence:hw1, wmm:fence:hw3, wmm:fence:hw4} due to the
inherent complexities involved caused by the intricate semantics of such
architectures and fences.  Thus, testing-based methods are of limited use in
detecting weak memory defects, which suggests that a more systematic
analysis is needed to locate these defects.

In this work, we present a novel, true-concurrency inspired investigation
that leverages symbolic Bounded Model Checking (BMC) to locate defects in
modern weak memory programs.  We begin by introducing the problem of
assertion checking in weak memory programs using pre-C11 programs as
exemplar, and introduce the concerns that motivate our approach.

\begin{figure}[t]
\centering
\begin{tabular}{cc}
\subcaptionbox{\label{fig:tso:reorder}}[.45\linewidth]
{
\centering
\begin{scriptsize}
\begin{tabular}{c@{}c@{}c}
\multicolumn{3}{c}{\MT{x=0,y=0;}}\\
& & \\
\begin{minipage}{0.175\linewidth}
$s_1$  :  \MT{x\phantom{1}=1;}\\
$s_2$  :  \MT{r1=y;}\\
\end{minipage} & \large{$\parallel$} &
\begin{minipage}{0.16\linewidth}
$s_3$  :  \MT{y\phantom{1}=1;}\\
$s_4$  :  \MT{r2=x;}\\
\end{minipage}\\
& & \\
\multicolumn{3}{c}{\MT{assert(r1==1\phantom{1}||\phantom{1}r2==1)}}\\
\end{tabular}
\end{scriptsize}
}
&
\subcaptionbox{\label{fig:pso:reorder}}[.45\linewidth]
{
\centering
\begin{scriptsize}
\begin{tabular}{c@{}c@{}c}
\multicolumn{3}{c}{$\MT{x=0,y=0;}$}\\
& & \\
\begin{minipage}{0.16\linewidth}
$s_1$:  \MT{x = 1;}\\
$s_2$:  \MT{y = 1;}\\
\end{minipage} & \large{$\parallel$} &
\begin{minipage}{0.16\linewidth}
$s_3$:  \MT{r1=y;}\\
$s_4$:  \MT{r2=x;}\\
\end{minipage}\\
& & \\
\multicolumn{3}{c}{\MT{assert(r1!=1\phantom{1}||\phantom{1} r2==1)}}\\
\end{tabular}
\end{scriptsize}
}
\end{tabular}
\caption{
(\subref{fig:tso:reorder})~Reordering in \tso~~~
(\subref{fig:pso:reorder})~Reordering in \pso}
\end{figure}

\subsection{Example}

Consider the program given in \cref{fig:tso:reorder}.  Let \TT{x} and \TT{y}
be shared variables that are initialised with 0.  Let the variables \TT{r1}
and \TT{r2} be thread local.  Statements $s_1$ and $s_3$ both perform write
operations.  However, owing to store-buffering, these writes may not be
immediately dispatched to the main memory.  Next, after performing $s_1$ and
$s_3$, both threads may now proceed to perform the read operations $s_2$ and
$s_4$.  Since the write operations might still not have hit the memory,
stale values for \TT{x} and \TT{y} may be read into \TT{r2} and \TT{r1},
respectively.  This may cause the assertion to fail.  Such a behaviour is
possible in a processor that implements \emph{Total Store Ordering}~(\tso),
which permits \emph{weakening} (or \emph{relaxing}) the write-read ordering when the
operations are performed on different memory locations. 
Note that on a hypothetical architecture that guarantees \seqc, this would
never happen.  However, due to store buffering, a global observer might
witness that the statements were executed in the order $s_2; s_4; s_1; s_3$
which resulted in the said assertion failure.  We say that the events inside
the pairs $\inroundb{s_1,s_2}$ and $\inroundb{s_3,s_4}$ have been
\defw{reordered}.

\cref{fig:pso:reorder} illustrates how the assertion might fail on
architectures that implement \emph{Partial Store Order}~(\pso), which
permits write-write and write-read reorderings when these operations are on
different memory locations.  If \seqc was honoured, one
would expect to observe $\MT{r2==1}$ if $\MT{r1==1}$ has been observed.
However, reordering of the write operations $(s_1,s_2)$ (under \pso) would
lead to the assertion failure.

In this work we would like to find assertion violations that occur in
programs written for modern multiprocessors.  Specifically, we will be
focussing on C programs written for architectures that implement
reordering-based memory models like \tso and \pso.  We assume that the
assertions to be checked are given as part of the program text.

\subsection{Our Approach}

Our approach differs from most existing research in the way we model
concurrency.  Most traditional work rely exclusively on interleaving
semantics to reason about real-world multiprocessor programs.  An
interleaving semantics purports to reduce concurrent computations to their
equivalent non-deterministic \emph{sequential} computations.  For instance,
let $P$ be a system with two concurrent events $a$ and $b$; let's denote
this fact as $P$ $\define$ $a\|b$.  Interleaving semantics then assigns the
following meaning to $P$: $\semantics{P}$ $\define$ $a.b$ $|$ $b.a$ where
`$.$' denotes sequential composition and `$|$' denotes non-deterministic
choice.  That is, a system in which $a$ and $b$ happens in parallel is
indistinguishable from a system where $a$ and $b$ could happen in any order;
we call $a.b$ (also $b.a$) a \defw{schedule}.  The set of all possible
schedules is called the \defw{schedule space} of $P$.  As the input program
size increases, the (interleaved) schedule space of
the program may grow exponentially.  This schedule space explosion severely
limits the scalability of many analyses.  The state space issues caused by
interleaving semantics are only exacerbated under weak memory systems:
weakening a memory model increases the number of admissible reorderings, and
with it the degree of non-determinism.

In this paper, we propose an alternative approach: to directly capture the
semantics of shared memory programs using \defw{true concurrency}~\cite{dimacs:pratt, conctheory:bowman-gomez}.
There are two competing frameworks for
constructing a truly concurrent semantics, one based on \defw{event
structures}~\cite{estruct:winskel86} and the other on
\defw{pomsets}~\cite{pomsets:pratt:1986}.  A recent paper~\cite{cav13}, the
work that is closest to ours, uses partial orders (pomsets) to capture the
semantics of shared memory program.  The main insight is that partial orders
neatly capture the causality of events in the dynamic execution of weak
memory programs.  But such a model cannot directly capture the control and
data flow choices present in the programs: the semantics of programs with
multiple, conflicting (that is, mutually exclusive) dynamic executions is
captured simply as a \emph{set of} candidate executions~\cite{cav13}.  In
this work, we advocate integrating program choices \emph{directly} into the
true concurrency semantics: we show that this results in a more succinct,
algebraic presentation, leading to a more efficient analysis.  Our true
concurrency semantics employs general event structures which, unlike partial
orders, tightly integrate the branching/conflict structure of the program
with its causality structure.  Intuitively, such a conflict-aware truly
concurrent meaning of $P$ can be given as follows: $\semantics{P}$ $\define$
$\neg (a\#b)$ $\wedge$ $\neg (a < b)$; that is, the events $a$ and $b$ are
said to be concurrent iff they are not conflicting (\#) and are not
related by a `happens-before' ($<$) relation.
This (logical) characterisation is strictly more expressive compared to
interleaving-based characterisation\footnote{For systems with a finite set
of events, the expressibility of both the notions of concurrency
coincides.}; in addition, such a semantics does not suffer from the schedule
space explosion problem of a (more operationally defined) interleaving
semantics.  Intuitively, a true concurrency admits the phenomenon of
concurrency to be a \emph{true}, a first-class phenomenon that exists in the
real-world and needed to modelled as such, as opposed to simply reducing it
to a sequentialised choice.

Our event structure based semantics can naturally distinguish computations
at a granularity finer than trace equivalence.  For instance, consider
threads $t_1$, $t_2$ and $s$, defined as follows: $t_1 = a.b$ $|$ $a.c$,
$t_2 = a.(b|c)$ and $s = a.c$.  Note than $t_1$ and $t_2$ are trace (and
partial order) equivalent.  Let $\|$ denote the (synchronous) parallel
composition~\cite{csp}.  Then, $t_1 \| s$ can deadlock while $t_2 \| s$
cannot.  Our semantics can distinguish $t1$ from $t2$, while current partial
order based methods cannot.  This is because partial order based methods
capture the semantics as a set of \emph{complete executions}, without ever
specifying how the partial sub-executions that constitute a trace unfold. 
Thus, our semantics can be used to reason about deadlocks over partial
computations involving, say, lock/unlock operations --- not just assertion
checking over complete computations involving read/write operations.

Although event structures offer an attractive means to capture the semantics
of weak memory programs, we are not aware of any event structure based tool
that verifies modern weak memory programs written using real-world languages
like~C/C++.  In this work, we address this lacuna: by investigating the
problem of assertion checking in weak memory programs using event
structures.  We first develop a novel true concurrency semantics based on
general event structures~\cite{estruct:winskel86} for a {\em bounded model
checker}.  We then formulate a succinct symbolic decision procedure and use
this decision procedure to locate assertion violations in modern shared
memory programs.  Our tool correctly and efficiently analyses the large,
representative set of programs in the SV-COMP~2015~\cite{sv-comp15}
benchmark suite.  It also successfully handles all the intricate tests in
the widely used Litmus suite~\cite{litmus}.

\subsection{Contributions}

Following are our contributions.
\begin{enumerate}

\item A compositional, symbolic (event structure based) true concurrency
semantics for concurrent programs, and a characterisation of assertion
violation over this semantics~(\cref{sec:sem}).

\item A novel decision procedure based on the above semantics that locates
assertion violations in multi-threaded programs written for weak memory
architectures~(\cref{sec:enc}).

\item A BMC-based tool that implements our ideas and a thorough performance
evaluation on real-world C programs of our approach against the
state-of-the-art research~(\cref{sec:eval}).

\end{enumerate}

We present our work in to three parts. The first part introduces the relevant
background on weak memory models~(\cref{sec:bkground}) and true
concurrency~(\cref{sec:bkground:tc}).  The second part defines an abstract,
true concurrency based semantics for weak memory programs written
in~C~(\cref{sec:sem}) and presents a novel decision procedure that exploits
this abstract semantics to find assertion violations~(\cref{sec:enc}).  In
the third part we discuss the specifics of our tool and present a thorough
performance evaluation~(\cref{sec:eval}); we then discuss the related work
in~\cref{sec:rel} and conclude.

\begin{figure*}[t]
\centering
\begin{tabular}{ccc}
\hspace{-3em}
\subcaptionbox{\label{sc:model}}[.33\linewidth]
{
\centering
\scriptsize
\begin{tikzpicture}
\node[draw=none] (procs) {\textsc{processors}};
\node[draw=none, below=.01ex of procs] (p1) {$P_1$};
\node[draw=none, left=3em of p1] (p0) {$P_0$};
\node[draw=none, right=3em of p1] (p2) {$P_2$};
\node[draw=none, below left=7ex and 0.01em  of p0.south] (r0) {r};
\node[draw=none, below right=5ex and 0.01em of p0.south] (w0) {w};
\node[draw=none, below left=7ex and 0.01em  of p1.south] (r1) {r};
\node[draw=none, below right=5ex and 0.01em of p1.south] (w1) {w};
\node[draw=none, below left=7ex and 0.01em  of p2.south] (r2) {r};
\node[draw=none, below right=5ex and 0.01em of p2.south] (w2) {w};
\node[box, below=.1ex of p0] (p0Box) {};
\node[box, below=.1ex of p1] (p1Box) {};
\node[box, below=.1ex of p2] (p2Box) {};
%
\node[draw=none, below=16ex of p1] (swt) {};
%
\node[circ, above=0.1ex of swt] (swtt) {};
\node[circ, below=0.1ex of swt] (swtb) {};
\node[circ, left=0.1ex  of swt] (swtl) {};
\node[circ, right=0.1ex of swt] (swtr) {};
\node[box, minimum height=1.1em,  below=0.85ex of swtb] (mem) {\phantom{2}\textsc{single port memory}\phantom{2}};
%
\draw [-]  (p0Box.south) -- (p0Box.south|-swtt.north);
\draw [->] (p0Box.south|-swtt.north) -- (swtl.west);
\draw [->] (p1Box.south) -- (p1Box.south|-swtt.north);
\draw [-]  (p2Box.south) -- (p2Box.south|-swtt.north);
\draw [->] (p2Box.south|-swtt.north) --  node [below=0.5ex,  right] {commits} (swtr.east);
\draw [-]  (swtb.south) -- (swtb.south|-mem.north);
\draw [-]  (swtb.north) -- ($(swtb.north|-swtt.south) + (0pt, -0.75pt)$) ;
\end{tikzpicture}
}
&
\subcaptionbox{\hspace{2em}\label{tso:model}}[.33\linewidth]
{
\centering
\scriptsize
\begin{tikzpicture}
\node[draw=none] (procs) {\textsc{processors}};
\node[draw=none, below=.01ex of procs] (p1) {$P_1$};
\node[draw=none, left=4em of p1] (p0) {$P_0$};
\node[draw=none, right=4em of p1] (p2) {$P_2$};
\node[draw=none, below left=7ex and 0.01em  of p0.south] (r0) {r};
\node[draw=none, below right=2ex and 0.5em  of p0.south] (w0) {w};
\node[draw=none, below left=7ex and 0.01em  of p1.south] (r1) {r};
\node[draw=none, below right=2ex and 0.5em  of p1.south] (w1) {w};
\node[draw=none, below left=7ex and 0.01em  of p2.south] (r2) {r};
\node[draw=none, below right=2ex and 0.5em  of p2.south] (w2) {w};
\node[box, below=.1ex of p0] (p0Box) {};
\node[box, below=.1ex of p1] (p1Box) {};
\node[box, below=.1ex of p2] (p2Box) {};
%
\node [tsobuffhead, above right=.1ex and 0.5em of r0] (head0) {};
\draw [-, draw=white, thick] (head0.north east) -- (head0.north west);
\node [tsobuff, anchor=north] at (head0.south)   (p0Buff0) {};
\node [tsobuff, anchor=north] at (p0Buff0.south) (p0Buff1) {};
\node [tsobuff, anchor=north] at (p0Buff1.south) (p0Buff2) {};
\node [tsobuff, anchor=north] at (p0Buff2.south) (p0Buff3) {};
%
\node [tsobuffhead, above right=.1ex and 0.5em of r1] (head1) {};
\draw [-, draw=white, thick] (head1.north east) -- (head1.north west);
\node [tsobuff, anchor=north] at (head1.south)   (p1Buff0) {};
\node [tsobuff, anchor=north] at (p1Buff0.south) (p1Buff1) {};
\node [tsobuff, anchor=north] at (p1Buff1.south) (p1Buff2) {};
\node [tsobuff, anchor=north] at (p1Buff2.south) (p1Buff3) {};
%
\node [tsobuffhead, above right=.1ex and 0.5em of r2] (head2) {};
\draw [-, draw=white, thick] (head2.north east) -- (head2.north west);
\node [tsobuff, anchor=north] at (head2.south)   (p2Buff0) {};
\node [tsobuff, anchor=north] at (p2Buff0.south) (p2Buff1) {};
\node [tsobuff, anchor=north] at (p2Buff1.south) (p2Buff2) {};
\node [tsobuff, anchor=north] at (p2Buff2.south) (p2Buff3) {};
\node [draw=none, right=0.1em of head2]        (p2LabelA) {FIFO};
\node [draw=none, below right=0.01ex and 0.1em of head2]   (p2LabelB) {store};
\node [draw=none, below right=2ex and 0.1em of head2]   (p2LabelC) {buffer};
\node[draw=none, below=16ex of p1] (swt) {};
%
\node[circ, above=0.1ex of swt] (swtt) {};
\node[circ, below=0.1ex of swt] (swtb) {};
\node[circ, left=0.1ex  of swt] (swtl) {};
\node[circ, right=0.1ex of swt] (swtr) {};
%
\node[box, minimum height=1.1em,  below=.85ex of swtb] (mem) {\phantom{2}\textsc{single port memory}\phantom{2}};
\draw [<-, name path=path0]  (p0Box.south) -- (p0Box.south|-swtt.north);
\draw [->] (p0Box.south|-swtt.north) -- (swtl.west);
\draw [->] ($ (p0Box.south) + (3pt,0pt) $) -- (head0.north);
\draw [-] (p0Buff3.south) -- ($ (p0Buff3.south) + (0pt, -2ex) $);
\path [-, name path=line0] ($ (p0Buff3.south) + (0pt, -2ex) $) -- ($ (p0Buff3.south) + (-1em, -2ex) $) ;
\draw [->, name intersections={of=line0 and path0, by={Int0}}] ($ (p0Buff3.south)  + (0pt, -2ex) $) -- (Int0);
\draw [<->, name path=path1] (p1Box.south) -- (p1Box.south|-swtt.north);
\draw [->] ($ (p1Box.south) + (3pt,0pt) $) -- (head1.north);
\draw [-] (p1Buff3.south) -- ($ (p1Buff3.south) + (0pt, -2ex) $);
\path [-, name path=line1] ($ (p1Buff3.south) + (0pt, -2ex) $) -- ($ (p1Buff3.south) + (-1em, -2ex) $) ;
\draw [->, name intersections={of=line1 and path1, by={Int0}}] ($ (p1Buff3.south)  + (0pt, -2ex) $) -- (Int0);
\draw [<-, name path=path2]  (p2Box.south) -- (p2Box.south|-swtt.north);
\draw [->] (p2Box.south|-swtt.north) --  node [below=0.5ex, right] {commits} (swtr.east);
\draw [->] ($ (p2Box.south) + (3pt,0pt) $) -- (head2.north);
\draw [-] (p2Buff3.south) -- ($ (p2Buff3.south) + (0pt, -2ex) $);
\path [-, name path=line2] ($ (p2Buff3.south) + (0pt, -2ex) $) -- ($ (p2Buff3.south) + (-1em, -2ex) $) ;
\draw [->, name intersections={of=line2 and path2, by={Int0}}] ($ (p2Buff3.south)  + (0pt, -2ex) $) -- (Int0);
\draw [-]  (swtb.south) -- (swtb.south|-mem.north);
\draw [-]  (swtb.north) -- ($(swtb.north|-swtt.south) + (0pt, -0.75pt)$) ;
\end{tikzpicture}
}
&
\hspace{-5em}
\subcaptionbox{\label{pso:model}}[.33\linewidth]
{
\centering
\scriptsize
\begin{tikzpicture}
\node[draw=none] (procs) {\textsc{processors}};
\node[draw=none, below=0.865ex of procs] (p1) {};
\node[draw=none, left=4em of p1] (p0) {$P_0$};
\node[draw=none, right=4em of p1] (p2) {$P_1$};
\node[draw=none, below left=7ex and 0.01em  of p0.south] (r0) {r};
\node[draw=none, below right=2ex and 0.75em of p0.south] (w0) {w};
\node[draw=none, below left=7ex and 0.01em  of p2.south] (r2) {r};
\node[draw=none, below right=2ex and 0.75em of p2.south] (w2) {w};
\node[box, below=.1ex of p0] (p0Box) {};
\node[box, below=.1ex of p2] (p2Box) {};
%
\node [draw=none, above right=1.25ex and 0.125em of r0, align=center] (varhead0) {\tiny x\phantom{1}...z};
\node [psobuffhead, above right=0.1ex and 0.5em of r0, align=center] (head0) {
\nodepart{one}
\nodepart{two}
\nodepart{three}
};
\draw [-, draw=white, thick] (head0.north east) -- (head0.north west);
\node [psobuff, anchor=north] at (head0.south)   (p0Buff0) {
\nodepart{one}
\nodepart{two}
\nodepart{three}
};
\node [psobuff, anchor=north] at (p0Buff0.south) (p0Buff1) {
\nodepart{one}
\nodepart{two}
\nodepart{three}
};
\node [psobuff, anchor=north] at (p0Buff1.south) (p0Buff2) {
\nodepart{one}
\nodepart{two}
\nodepart{three}
};
\node [psobuff, anchor=north] at (p0Buff2.south) (p0Buff3) {
\nodepart{one}
\nodepart{two}
\nodepart{three}
};
%
\node [draw=none, above right=1.25ex and 0.125em of r2, align=center] (varhead2) {\tiny x\phantom{1}...z};
\node [psobuffhead, above right=.1ex and 0.5em of r2, align=center] (head2)  {
\nodepart{one}
\nodepart{two}
\nodepart{three}
};
;
\draw [-, draw=white, thick] (head2.north east) -- (head2.north west);
\node [psobuff, anchor=north] at (head2.south)   (p2Buff0)  {
\nodepart{one}
\nodepart{two}   
\nodepart{three}
};
\node [psobuff, anchor=north] at (p2Buff0.south) (p2Buff1)  {
\nodepart{one}
\nodepart{two}   
\nodepart{three}
};
\node [psobuff, anchor=north] at (p2Buff1.south) (p2Buff2)  {
\nodepart{one}
\nodepart{two}   
\nodepart{three}
};
\node [psobuff, anchor=north] at (p2Buff2.south) (p2Buff3) {
\nodepart{one}
\nodepart{two}   
\nodepart{three}
};
\node [draw=none, right=0.1em of head2]        (p2LabelA) {per-addr};
\node [draw=none, below right=0.01ex and 0.1em of head2]   (p2LabelB) {store};
\node [draw=none, below right=2ex and 0.1em of head2]   (p2LabelC) {buffer};
\node[draw=none, below=17ex of p1] (swt) {};
%
\node[draw=none, above=0.1ex of swt] (swtt) {};
\node[circ, below=0.1ex of swt] (swtb) {};
\node[circ, left=0.1ex  of swt] (swtl) {};
\node[circ, right=0.1ex of swt] (swtr) {};
%
\node[box, minimum height=1.1em,  below=.85ex of swtb] (mem) {\phantom{2}\textsc{single port memory}\phantom{2}};
\draw [<-, name path=path0]  (p0Box.south) -- (p0Box.south|-swtt.north);
\draw [->] (p0Box.south|-swtt.north) -- (swtl.west);
\draw [->] ($ (p0Box.south) + (3pt,0pt) $) -- ($ (head0.north) $);
\draw [-] (p0Buff3.south) -- ($ (p0Buff3.south) + (0pt, -2ex) $);
\path [-, name path=line0] ($ (p0Buff3.south) + (0pt, -2ex) $) -- ($ (p0Buff3.south) + (-5em, -2ex) $) ;
\draw [->, name intersections={of=line0 and path0, by={Int0}}] ($ (p0Buff3.south)  + (0pt, -2ex) $) -- (Int0);
\path [<-, name path=path2]  (p2Box.south) -- (p2Box.south|-swtt.north);
\draw [<-]  (p2Box.south) -- (p2Box.south|-swtt.north);
\draw [->] (p2Box.south|-swtt.north) --  node [below=0.5ex, right] {commits} (swtr.east);
\draw [->] ($ (p2Box.south) + (3pt,0pt) $) -- (head2.north);
\draw [-] (p2Buff3.south) -- ($ (p2Buff3.south) + (0pt, -2ex) $);
\path [-, name path=line2] ($ (p2Buff3.south) + (0pt, -2ex) $) -- ($ (p2Buff3.south) + (-5em, -2ex) $) ;
\draw [->, name intersections={of=path2 and line2, by={Int0}}] ($ (p2Buff3.south)  + (0pt, -2ex) $) -- (Int0);
\draw [-]  (swtb.south) -- (swtb.south|-mem.north);
\draw [-]  (swtb.north) -- ($(swtr.west) + (0, -0.75pt)$) ;
\end{tikzpicture}
}
\end{tabular}
\caption{
Execution Models: 
(\subref{sc:model}) SC\,\,
(\subref{tso:model}) TSO\,\,
(\subref{pso:model}) PSO}
\label{fig:models}
\end{figure*}
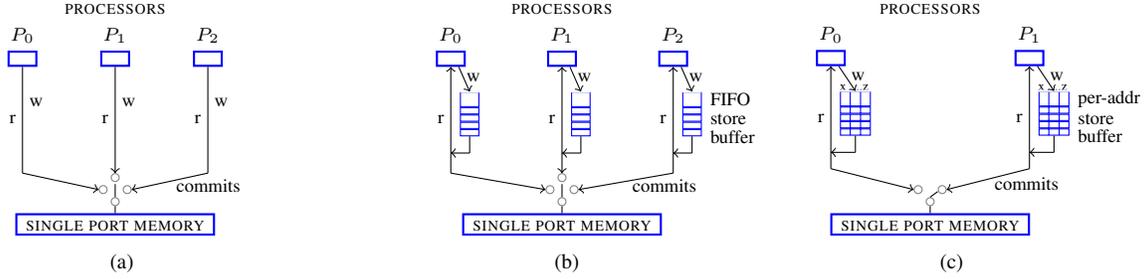

\section{Background}
\label{sec:bkground}

This section summarises the ideas behind three concepts: weak memory models,
bounded model checking, and our intermediate program representation.

\subsection{Weak Memory Models}
\label{sec:bkground:wmm}

We introduce three memory models --- \seqc, \tso and \pso{} --- and the
necessary intuitions to understand them.  We currently do not directly
handle other memory models; please consult the related work section for
further discussion on this.

\hfill\\\textsc{Sequential Consistency} (\seqc):
This is the simplest and the most intuitive memory model where executions
strictly maintain the intra-thread program order (\progo), while permitting
arbitrary interleaving of instructions from other threads.  Intuitively, one
could view the processors/memory system that offers SC as a single-port
memory system where the memory port is connected to a switch/multiplexer,
which is then connected to the processors.  This switch can only commit one
instruction from a thread/processor at a time, and it does so
non-deterministically; this is depicted in~\cref{sc:model}.  The
illustrations in~\cref{fig:models} are from the SPARC architecture
manual~\cite{sparc:manual:8}.

\hfill\\\textsc{Total Store Order}~(\tso):
In \tso, in addition to the behaviours permitted in \seqc, a write followed
by a read to a different memory location may be reordered.  Thus, the set of
executions permissible under \tso is a strict superset of \seqc.  This
memory model is used in the widely deployed x86 architecture.  Writes in x86
are first enqueued into the \emph{store buffer}.  These writes are later
committed to memory in the order in which they are issued, i.e., the program
order of writes is preserved.  The store buffer also serves as a read cache
for the future reads from the same processor.  But any read to a variable
that is not in the store buffer can be issued directly to memory and such
reads could be completed before the pending, enqueued writes hit the memory. 
Well known mutual exclusion algorithms like Dekker, Peterson and Parker are
all unsafe on \/x86.  A~\tso processor/memory system could be seen as one
where the processor issues writes to a store buffer, but sends the reads
directly to memory; this is depicted in~\cref{tso:model}.

\hfill\\\textsc{Partial Store Order}~(\pso):
\pso is \tso with an additional relaxation: \pso guarantees that only writes
to the \emph{same location} are committed in the order in which they are
issued, whereas writes to different memory locations may be committed to
memory out of order.  This is intuitively captured by a processor/memory
system that employs separate store buffers for writes that write to
different memory addresses, but the reads are still issued directly to the
memory; this is depicted in~\cref{pso:model}.

\subsection{Bounded Model Checking}
\label{sec:bkground:bmc}

Bounded Model Checking~(\bmc) is a Model Checking technique that performs a
depth-bounded exploration of the state space.  This depth is often given in
the form of an unwinding limit for the loops in the program.  Bugs that
require longer paths (deeper unwindings) are missed, but any bug found is
indeed a real bug.  Bounded Model Checkers typically employ SAT/SMT-based
symbolic methods to explore the program behaviour exhaustively up to the
given depth.  As modern SAT solvers are able to solve propositional formulas
that contain millions of variables, \bmc is increasingly being used to
verify real-world programs~\cite{bmc:sat}.

\begin{lrbox}{\eglstbox}
\centering
\begin{minipage}{0.45\linewidth}
\begin{flushleft}
Thread 1:
\vspace{-2ex}
\end{flushleft}
\centering
\begin{lstlisting}[numbers=none, language=C]
x = 1;       // Wx3
if(y % 2)    // Ry3
   x = 3;    // Wx4
else 
   x = 7;    // Wx5
y = x;       // Wy4, Rx6
\end{lstlisting}
\end{minipage}
\hspace{2em}
\begin{minipage}{0.45\linewidth}
\begin{flushleft}
Thread 2:
\vspace{-2ex}
\end{flushleft}
\centering
\begin{lstlisting}[numbers=none, language=C]
y = 1;       // Wy5
if(x % 2)    // Rx7
   y = 3;    // Wy6
else 
   y = 7;    // Wy7
x = y;       // Wx8, Ry8
\end{lstlisting}
\end{minipage}
\end{lrbox}
\begin{lrbox}{\egguardbox}
\centering
\begin{minipage}{0.45\linewidth}
\vspace{-2ex}
\centering
\begin{lstlisting}[numbers=none, mathescape=true]
guard$_1 \define \top$
guard$_1 \Rightarrow $(x$_3$=1)
guard$_2 \define $(y$_3$%2=0)
guard$_3 \define $(guard$_1 \wedge $guard$_2$)
guard$_4 \define $(guard$_1 \wedge \neg$guard$_2$)
guard$_4 \Rightarrow $(x$_4$=3)
guard$_3 \Rightarrow $(x$_5$=7)
guard$_1 \Rightarrow $(x$_{\phi1}$=guard$_2$?x$_5$:x$_4$)
guard$_1 \Rightarrow $(y$_4$=x$_6$)
\end{lstlisting}
\end{minipage}
\hspace{2em}
\begin{minipage}{0.45\linewidth}
\centering
\vspace{-2ex}
\begin{lstlisting}[numbers=none, mathescape=true]
guard$_5 \define \top$ 
guard$_5 \Rightarrow $(y$_5$=1)
guard$_6 \define $(x$_7$%2=0)
guard$_7 \define $(guard$_5 \wedge $guard$_6$)
guard$_8 \define $(guard$_5 \wedge \neg$guard$_6$)
guard$_8 \Rightarrow $(y$_6$=3)
guard$_7 \Rightarrow $(y$_7$=7)
guard$_5 \Rightarrow $(y$_{\phi1}$=guard$_6$?y$_7$:y$_6$)
guard$_5 \Rightarrow $(x$_8$=y$_8$)
\end{lstlisting}

\end{minipage}

\end{lrbox}

\begin{figure*}%
\centering
\begin{tabular}{cc}
\begin{subfigure}{0.55\linewidth}
\centering
\usebox{\eglstbox}
\usebox{\egguardbox}
\vspace{-2ex}
\caption{}
\label{fig:wmm-bmc:eg:pgm}
\end{subfigure}
&
\hspace{-6em}
\begin{subfigure}{0.55\linewidth}
\scriptsize
\centering
\tikzset{
    punkt/.style={
           rectangle,
			fill=white,
           draw=black, 
           text width=1.25em,
           minimum height=0.25ex,
           text centered},
    pil/.style={
           ->,
           shorten <=2pt,
           shorten >=2pt,}
}
\begin{tikzpicture}[->,>=stealth',shorten >=1pt,auto,node distance=4.5em,
                   remember picture]
  \node[draw=none]	(00) 		   		{};
  \node[punkt]  (10) [above right=1.5em and 1em of 00]	{\tt Ry$_3$};  
  \node[punkt]  (11) [above right=5em and 1em of 00]	{\tt Wx$_3$};  
  \node[punkt]  (12) [below right=0.25em and 2.25em of 11] {\tt Wx$_4$};
  \node[punkt]  (13) [right of=12] {\tt Wx$_5$};
  \node[punkt]  (14) [right of=13] {\tt Rx$_6$};
  \node[punkt]  (15) [right of=14] {\tt Wy$_4$};
  \node[draw=none] (16) [above right =0.5em and 0em of 15] {Thread 1};
  \node[punkt]  (20) [below right=1.5em and 1em of 00] {\tt Rx$_7$};  
  \node[punkt]  (21) [below right=5em and 1em of 00] {\tt Wy$_5$};  
  \node[punkt]  (22) [above right=0.25em and 3em of 21] {\tt Wy$_6$};
  \node[punkt]  (23) [right of=22] {\tt Wy$_7$};
  \node[punkt]  (24) [right of=23] {\tt Ry$_8$};
  \node[punkt]  (25) [right of=24] {\tt Wx$_8$};
  \node[draw=none] (26) [below right =0.25em and 0em of 25] {Thread 2};
  \path (10) edge    node {} (12)
  		(11) edge    node {} (12)
        (12) edge    node {} (13)
        (13) edge    node {} (14)
        (14) edge    node {} (15)
        
        (20) edge    node {} (22)
        (21) edge    node {} (22)
        (22) edge    node {} (23)
        (23) edge    node {} (24)
        (24) edge    node {} (25);
%
        \draw[-,dashed, magenta] (14.south) -- (25.north);
        \draw[-,dashed, bend right, magenta] (14.north) to (12.north);
        \draw[-,dashed, bend right, magenta] (14.north) to (13.north);
        \draw[-,dashed, bend right=30, blue] (10.south west) to (21.north west);
        \draw[-,dashed, blue] (10) -- (22);
        \draw[-,dashed, blue] (10) -- (23);

        \draw[-,dashed, blue] (24.north) -- (15.south);
        \draw[-,dashed, bend left, blue] (24.south) to (22.south);
        \draw[-,dashed, bend left, blue] (24.south) to (23.south);
        \draw[-,dashed, bend left=30, magenta] (20.north west) to (11.south west);
        \draw[-,dashed, magenta] (20) -- (12);
        \draw[-,dashed, magenta] (20) -- (13);
		
\draw[-, loosely dashdotted, black] (-0.25,0) to (6.25, 0);
\end{tikzpicture}
\vspace{15ex}
\caption{}
\label{fig:wmm-bmc:eg:ppo-potmat}
\end{subfigure}
\end{tabular}
\vspace{-1ex}
\caption{A program and its corresponding intermediate form}
\label{fig:wmm-bmc:eg}
\end{figure*}
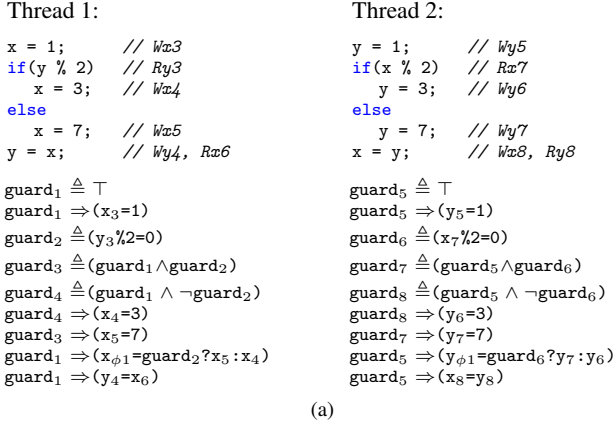
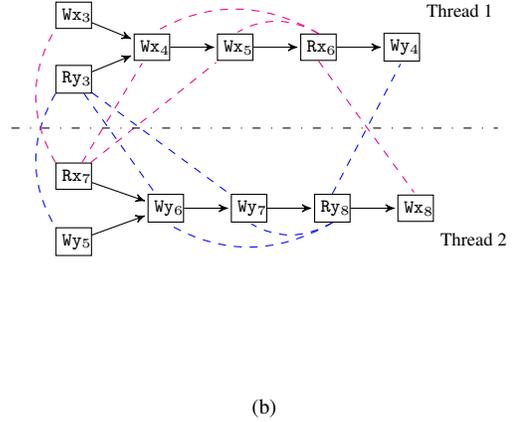

\subsection{Program Representation \label{sec:pr}}

We rely on a Bounded Model Checker that uses a symbolic static single
assignment form~(\ssa) to represent the input program.  Specifically, we use
\cbmc~\cite{cbmc}, a well-known bounded model checker that supports C99 and
most of the C11 standard.  It can handle most compiler extensions provided
by gcc and Visual Studio.  Over this \cbmc-generated symbolic \ssa, we
define two additional relations: a per-thread binary relation called
\emph{preserved program order}~(\ppo) and a system-wide $n$-ary relation
called \emph{potential matches}~(\potmat), where $n$ is the total number of
threads in the system.  The triple ---~\ssa, \ppo, and \potmat~--- is then
used to define a truly concurrent semantics of the input program; one can
see this triplet as an intermediate form that we use to represent all
relevant aspects of the input program.  These three components are discussed
in more detail below.

\hfill\\\textsc{Static Single Assignment Form} (\ssa):
The control and data flow of the input program is captured using guarded
\ssa form~\cite{cbmc:ssa}.  In a traditional SSA~\cite{compilers:ssa} or a
concurrent SSA~\cite{concurrent-ssa}, a distict symbolic variable is
introduced when a program variable appears on the left hand side (lhs) of an
assignment, whereas $\phi$ and $\pi$ function represents the possible values
that may flow into an assignment.  In the guarded \ssa, each occurrence of
the shared variable is treated as a distinct symbolic variable, and is given
a unique \ssa index, essentially allowing right hand side (rhs) symbols to
remain unconstrained.  In the guarded \ssa, assignments are converted into
equalities and conditionals act as guards to enforce these equalities.  To
restrict the values to only those as permitted by the underlying memory
model, additional equality constraints specific to the memory model are then
added.  These constraints capture all possible interleavings and reordering
of read and write operations as required by the model.  Thus, the $\pi$
functions of concurrent \ssa are subsumed by these constraints.  The details
of the encoding are provided in Section~\ref{sec:enc}.

We rely on the underlying \bmc tool to supply the necessary (symbolic)
variables and the constraints to cover the C constructs used in the input
program.  Each SSA assignment is decomposed into a pair of read and write
events.  These read/write events are augmented with guards over symbolic
program variables: this guard is a disjunction (induced by path merging) of
all conjunctions (induced by nested branch) of all paths that lead to the
read/write events.  For each uninitialised variable, we add a \defw{initial
write} that sets the variable to a non-deterministic, symbolic value.  From
now on, we will refer to our guarded \ssa simply as \ssa.

\hfill\\\textsc{Preserved Program Order} (\ppo):
\label{sec:wmem:ppo}
This is a per-thread binary relation specific to the mem\-ory-model that is
directed, irreflexive and acyclic.  \ppo captures the intra-thread ordering
of read/write events.  Given an input program in \ssa form, different memory
models produce different \ppo{}s.  Let \tppo be a binary relation over the
read/write events where, for every event $e_1$ and $e_2$, $\tuple{e_1, e_2}
\in \tppo$ iff the event $e_1$ cannot be relaxed after $e_2$.  Note that
\tppo is a partial order: it is transitive, anti-symmetric and reflexive. 
\tppo is collectively determined by the memory model under investigation and
the fences present in the input program.  We define \ppo to be the (unique)
transitive reduction~\cite{transreduct:72} of the \tppo.

\hfill\\\textsc{Potential Matches relation} (\potmat):
\label{sec:wmem:potmat}
While \ppo models the intra-thread control and data flow, the potential
matches relation aims at the inter-thread data flow.  It is an
$n$-ary relation with two kinds of tuples.  Let $m$ be a tuple and let
$m(i)$ denote the $i^{th}$ entry (for thread~$i$) in the tuple.  The
first kind of tuple with one event --- where $m(i) = e$ --- captures the
idea that the event~$e$ (in thread $i$) can happen by itself; the remaining
tuple entries contain `*'.  We say that such an $e$ is a \defw{free
event}.  Note that writes are free events, as they can happen by themselves. 

Let $i$ and $j$ be two \emph{distinct} thread indices that is, $i \ne j$ and
--- $0 \le i, j < n$.  The second kind of tuple,
involving two events --- where $m(i) = r$ and $m(j) = w$ --- denotes a
\emph{potential} inter-thread communication where a read $r$~(from thread
$i$) has read the value written by the write $w$~(from thread $j$); the rest
of the tuple entries contain `*'.  Such an $r$ is called a
\defw{synchronisation event}.  Reads are synchronisation events as they
cannot happen by themselves: reads always happen in conjunction with a free
(write) event.  One should see synchronisation events as events that
\emph{consume} other events, thus always needing another (free) event to
happen.  As we will see later~(\cref{sec:tc:ges:ops}), these two kinds of
tuples/events are fundamental to our semantics.

Informally, \potmat should be seen as a over-approximation of all possible
inter-thread choices/non-determinism available to each shared read (and
write) in a shared-memory program.  We assume that for every shared read,
there is at least one corresponding tuple ($m$) that \emph{matches} the said
read with a write: this corresponds to our intuition that every successful
read must have read from \emph{some} write.  We do not demand the converse:
there indeed could be writes that were not read by any of the reads.  Also, any
such $m$ can only relate reads and writes that operate on the \emph{same}
memory location: that is, reads and write are related by the potential matches 
iff they operate on the same (original) program variable.  A tuple in \potmat is
a potential inter-thread \defw{match} --- either containing an event that
matches with itself, or a pair of events that could match with one another --- 
hence the name potential matches.  We sometimes denote the potential matches
relation as $\potmatm$.  We currently construct~\potmatm as a (minimally
pruned) subset of the Cartesian product between per-thread events that share
the same address\footnote{More formally, \potmatm is proper subset of the
$n$-ary fibred product between per-thread event sets (say, $\esevents_i
\cup$ `*') where the event labels agree. The label of `*' agrees with all
the events in the system.}.  This subset consists only of the two
aforementioned types of tuples, and has at least one tuple for every read
(containing two events) and write (containing one event).

\hfill\\\textsc{Example}:
Consider the two-thread program in~\cref{fig:wmm-bmc:eg:pgm}. The bottom
half gives the corresponding \ssa form.  Both \pgmtxt{x} and \pgmtxt{y} are
shared variables.  The guards associated with each event can also be seen in
the figure.  As the distinct symbolic variables are introduced for every
occurrance of a program variable, assignments can be converted to guarded
equalities.  Guards capture the condition that must hold true to enforce an
equality. The symbols \texttt{guard$_3$} and \texttt{guard$_4$} illustrate how path
conditions are conjucted as we go along branches.  These symbols are also
said to guard the events participating in the equality.  For example,
\texttt{guard$_1 \Rightarrow$ (y$_4$=x$_6$)} denotes not only that
\texttt{guard$_1$} implies the equality but also that it acts as guard to
corresponding events $\pgmtxt{Wy}_4$ and $\pgmtxt{Rx}_6$: that is,
\pathcond($Wy_4$) = \pathcond($Rx_6$) = \texttt{guard$_1$}.  When the local
paths merge, auxiliary variables (e.g., \texttt{x$_{\phi1}$}) are
introduced, which hold appropriate intra-thread values depending upon which
path got executed.  Note that \texttt{x6} is completely free in the
constraints given in the figure.  Later on, additional constraints are
added, which restrict the value of \texttt{x6} to either an intra-thread
value of \texttt{x$_{\phi1}$} or an inter-thread value of \texttt{x$_8$}. 
The corresponding \tso intermediate form is given
in~\cref{fig:wmm-bmc:eg:ppo-potmat}: note that \tso relaxes the program
order between $\tuple{\pgmtxt{Wx}_3, \pgmtxt{Ry}_3}$ and
$\tuple{\pgmtxt{Wy}_5, \pgmtxt{Rx}_7}$.  The (intra-thread) solid arrows
depict the (intra-thread) preserved program order, and dashed lines depict
the potential matches relation.  The magenta lines show the matches involving
\pgmtxt{x} and the blue lines show the matches involving \pgmtxt{y}. The initial writes are omited for brevity.  The horizontal dash-dotted line demarcates the thread boundaries.


\ifx 0
\begin{lrbox}{\lstbox}
\centering
\begin{minipage}{0.45\linewidth}
\centering
\renewcommand{\lstlistingname}{Thread}
\begin{lstlisting}[numbers=none, language=C, caption={\hspace{1em}}]
// reads 42   
r1 = x;
if (r1 == 42)
  y = r1;
\end{lstlisting}
\vspace{4ex}
\end{minipage}
\begin{minipage}{0.45\linewidth}
\centering
\renewcommand{\lstlistingname}{Thread}
\begin{lstlisting}[numbers=none, language=C, caption={\hspace{1em}}]
// reads 42
r2 = y;
if (r2 == 42)
  x = 42;
else
  x = 42;
\end{lstlisting}
\end{minipage}
\end{lrbox}

\begin{figure*}%
\centering
\begin{subfigure}{0.45\linewidth}
\centering
\usebox{\lstbox}
\caption{A Simple C Program}
\label{fig:wmm-bmc:eg:pgm}
\end{subfigure}
\begin{subfigure}{0.45\linewidth}
\caption{Its Intermediate Form}
\label{fig:wmm-bmc:eg:ppo-potmat}
\end{subfigure}
\caption{A program and its corresponding intermediate form\label{figuretrace}}
\end{figure*}
\fi

\ifx 0
 with additional inter-thread edges, the
potential matches or the spawn edges.  There are three threads in this
program.  Thread $0$, first executes two spawn events $S_1$ and $S_2$ that
creates two more threads.  The dashed (blue) lines show the spawn edges that
connect a spawn event to the `start event' in the thread it created.  A
spawn event can seen as `synchronising' with the start event in the new
thread.  The
\MT{J} node is a join `event'.  Note that our modified \ssa assigns unique
ids to \emph{each} occurrence of a shared variable: for instance, $x_0$,
$x_1$, $x_2$ all account for $x$.  \ganeshP{Because the extra variables
introduce extraneous state space that needs to be 'constrained'?: We enforce
additional equality constrains (like $x_2 = x_0 \wedge x_2 = x_1$) to
suitably constraint the state space.  Solid edges depict the preserved
program order (here, under \tso).  The dotted lines depict the potential
matches relation.  Spawn edges are to be seen as a special kind of potential
matches edges.}

\fi

\section{True Concurrency}
\label{sec:bkground:tc}

Although most of the existing literature on event structures deal with prime
or stable event structures~\cite{estruct:winskel86}, we will be using (a
heavily modified) general event structure.  General event structures are
(strictly) more expressive compared to prime/stable event
structures~\cite{bundle:es, estruct:ccs}.  In addition, the constructions we
employ --- parallel composition and restriction of event structures --- have a
considerably less complex presentation over general event structures.

We now present the concepts and definitions related to event structures. In
each case, we give the formal definitions first, followed by an informal
discussion as to what these definitions capture.  Also, hereafter we will
simply say `event structures' to mean the modified general event structure
defined by us.

\hfill\\
\noindent A \textsc{General event structure} is a quadruple
$\gestruct{\esevents, \esconf, \enables, \eslfn}$\footnote{Hereafter, for
brevity, \gestruct{\esevents, \esconf, \enables} will stand for
\gestruct{\esevents, \esconf, \enables, \eslfn}: that is, every event
structure is implicitly assumed to be equipped with a label set $\Sigma$ and
a labling function $\eslfn: E \to \Sigma$.}, where:
\begin{itemize}
\renewcommand{\labelitemi}{$\bullet$}
\item $\esevents$ is a countable set of \defw{events}.

\item $\esconf \define \{X | X \subseteq_\mathit{finite} \esevents, \forall e_{1} \ne e_{2} \in X \Rightarrow (e_{1}, e_{2}) \notin \conflict\}$. $\conflict$ is an irreflexive, symmetric relation on $\esevents$, called the \defw{conflict relation}. Intuitively, $\esconf$ can be viewed as a collection of mutually consistent sets of events.

\item $\enables$ $\subseteq$ $\esconf \times \esevents$ is an \defw{enabling relation}.

\item $\eslfn : \esevents \to \eslset$ is a labeling function and $\eslset$
is a set of labels.

\end{itemize}
such that:
\begin{itemize}
\renewcommand{\labelitemi}{$-$}
\item $\esconf$ is consistent: $\forall X,Y \subseteq \esevents$, $X \subseteq Y$, $Y \in \esconf$ $\Rightarrow X \in \esconf$

\item $\enables$ is extensive: $\forall e \in \esevents$, $\forall X, Y \in \esconf$, $X \enables e$, $X \subseteq Y$ $\Rightarrow Y \enables e$

\end{itemize}

Let us now deconstruct the definition above. We would like to think of a
thread as a countable set of events ($\esevents$), which get executed in a
particular fashion.  Since we are interested only in finite computations, we
require that all execution `fragments' are finite.  Additionally, for
fragments involving conflicting events, we require that at most one of the
conflicting events occurs in the execution fragment.  The notion of
computational conflict (or choice) is captured by the conflict relation
(\#).  We call executions that abide by all the requirements above
\emph{consistent executions}; $\esconf$ denotes the set of all such consistent
executions.  Thus, $\esconf \subseteq 2^{\esevents}$ is the set of
conflict-free, finite subsets of $\esevents$.  Since we want the `prefixes'
of executions to be executions themselves, we demand that $\esconf$ is
subset closed.  Such execution fragments can be `connected' to events using
the enabling ($\enables$) relation: $X \enables e $ means that events of $X$
enable the event $e$.

For example, in an \seqc architecture if there is a write event $w$ followed
by a read event $r$, then $\tuple{\set{w}, r} \in$ $\enables$ as $w$ must
happen before $r$ could happen. In general, $\enables$ allows us to capture 
the dependencies within events as dictated by the underlying memory model. 
Note that since the enabling relation connects the elements of $\esconf$ with 
that of $\esevents$, it is automatically branching/conflict aware. 
We do not require that a set $X$  enabling $e$ to be the minimal set 
(enabling $e$): extensiveness only requires that $X$ contains a subset 
that enables $e$.  The labeling function, $\eslfn(e)$, returns the label 
of the read/write event $e$.  These labels are interpreted as addresses 
of the events. Finally, it is often useful to see $\esevents$ as a union 
of three disjoint sets $\esread$, $\eswrite$ and $\ireads$, where $\esread$ 
corresponds to the set of reads, $\eswrite$ to the set of writes and 
$\ireads$ correspond to the set of \defw{local reads}~(see~\cref{sec:sem}).

\hfill\\
\noindent\textsc{Configuration}: 
\label{sec:estruct:configuration}
A \defw{configuration} of event structure $\gestruct{\esevents, \esconf, \enables}$ is a subset $C \subseteq \esevents$ such that:
\begin{itemize}
\renewcommand{\labelitemi}{$-$}
\item $C$ is conflict-free: $C \in \esconf$

\item  $C$ is secured: $\forall e \in C$, $\exists e_{0}, \ldots, e_{n} \in C$, $e_{n} =
e$ $\wedge$ $\forall i$ $0 \le i \le n$ . $\setof{e_{0}, \ldots, e_{i-1}}  \enables e_{i}$
\end{itemize}

A configuration $C \subseteq \esevents$ is to be understood as a  history of
computation \emph{up to} some computational state.  This computational
history cannot include conflicting events, thus we would like all finite
subsets of $C$ to be conflict free; this can also be ensured by requiring
that all finite subsets of $C$ be elements of $\esconf$.  Securedness
ensures that for any event $e$ in a configuration, the configuration has as
subsets a sequence of configurations $\emptyset, \setof{e_{0}}, \ldots,
\setof{e_{0},\ldots,e_{n}}$ --- called a securing for $e$ in $C$, such that
one can build a `chain of enablings' that will eventually enable $e$; all
such chains must start from $\emptyset$.

Let the set of all configurations of the event structure
$\gestruct{\esevents, \esconf, \enables}$ be denoted by
\esconfigs{\esevents}.  A \defw{maximal configuration} is a configuration
that cannot be extended further by adding more events.

\hfill\\\noindent\textsc{Coincidence free}: 
\label{sec:estruct:coincidence}
Given an event structure $\gestruct{\esevents, \esconf, \enables}$, we say
that it is \defw{coincidence free} iff $\forall X \in
\esconfigs{\esevents}$, $\forall e, e' \in X$, $e \ne e' \Rightarrow \exists
Y \in \esconfigs{\esevents}$, $Y \subseteq X$, $(e \in Y \Leftrightarrow e'
\notin Y)$.

Intuitively, this property ensures that configurations add at most one event
at a time: this in turn ensures that secured configurations track the
enabling relation faithfully.  We require our event structures to be
coincidence free.  This is a technical requirement that enables us to assign
every event in a configuration a unique clock order (see below).

\hfill\\\noindent\textsc{Trace and Clock orders}:
Given an event $e$ in configuration~$C$ and a securing up to $e_k$ --- that
is, $\setof{e_{i=0},e_{i=1}, \ldots, e_{i=k-1}} \enables e_{i=k}$ --- we
define the following injective map $\securing{e}{C}: \restr{\esevents}{C}
\to \nats^{0}$ as $\securing{e}{C} \define i$.  Informally,
$\securing{e}{C}$ is the trace position of event $e$ in $C$:
$\securing{e_1}{C} < \securing{e_2}{C}$ implies that the event $e_1$ 
occurred before $e_2$ in the given securing of the configuration~$C$. 
Given such a $\secrng_C$ map, we define a monotone map named $\clk$ as 
$\clock{e}{C}: e \to \nats^{0}$ that is \emph{consistent} with $\secrng_C$. 
That is, $\forall e_1, e_2 \in C$, $\securing{e_1}{C} < \securing{e_2}{C} 
\Rightarrow \clock{e_1}{C} < \clock{e_2}{C}$.  Informally, the $\clk_C$ map 
\emph{relaxes} the $\secrng_C$ map monotonically so that $\clk_C$ can 
accommodate events from other threads, while still respecting the ordering 
dictated by $\secrng_C$.

\hfill\\\noindent\textsc{Partial Functions}:
As part of our event structure machinery, we will be working with partial
functions on events, say $f:\esevents_0 \to \esevents_1$.  The fact that $f$
is undefined for a $e \in \esevents_0$ is denoted by $f(e) = \bot$.  As a
notational shorthand, we assume that whenever $f(e)$ is used, it is
indeed defined.  For instance, statements like $f(e) = f(e')$ are
always made in a context where both $f(e)$ and $f(e')$ are indeed
defined.  Also, for a $X \subseteq \esevents_0$, $f(X) = \setof{f(e) \:|\:e
\in X \textnormal{ and } f(e)\textnormal{ is defined}}$.

\hfill\\\noindent\textsc{Morphisms}: 
A morphism between event structures is a structure-preserving function from
one event structure to another.  Let $\Gamma_{0} = \gestruct{\esevents_0,
\esconf_0, \enables_0}$ and $\Gamma_{1} = \gestruct{\esevents_1, \esconf_1,
\enables_1}$ be two stable event structures.

A \defw{ partially synchronous morphism}  $f:\Gamma_{0} \to \Gamma_{1}$ is a function $f$ from read set ($\esread_0$) to write set ($\eswrite_1$) such that:

\begin{itemize}
\renewcommand{\labelitemi}{$-$}
\item $f$ preserves consistency: $\forall X \in \esconf_{0}$ $\Rightarrow$ $f(X) \in \esconf_{1}$.

\item $f$ preserves enabling: $\forall X \enables_{0} e$,
$\defined(f(e))\footnote{The $\defined(f(e))$ predicate returns true if
$f(e)$ is defined.}$ $\Rightarrow$ $f(X) \enables_{1} f(e)$

\item $f$ preserves the labels: $f(e) = e'$ $\Rightarrow$ $\eslfn(e) = \eslfn(e')$

\item $f$ does not time travel: $X \in \esconf_0, Y\in \esconf_1$,  $f(e) = e'$ $\Rightarrow$ $\clock{e}{X} > \clock{e'}{Y}$

\item $f$ ensures freshness: $X \in \esconf_0, Y \in \esconf_1$, $f(e) = e'$, then $\forall e'' \in Y$ such that $\eslfn(e'') = \eslfn(e)$,  $\clock{e''}{Y} < \clock{e}{X} \Rightarrow \clock{e''}{Y} < \clock{e'}{Y}$
\end{itemize}
Such an $f$ is called \defw{synchronous morphism} if it is total.

A morphism should be seen as a way of synchronising reads of one event
structure with the writes of another.  We naturally require such a morphisms
to be a function in the set theoretic sense: this ensures that a read always
reads from exactly one write.  Note that the requirement of $f$ \defw{being
a function} introduces an implicit conflict between competing writes.  Given
a morphism $f:\Gamma_0 \to \Gamma_1$, $f(r_0)= w_1$ is to be understood as
$r_0$ reading the value written by $w_1$ (or $r_0$ \emph{synchronising} with
$w_1$).  Thus, the requirement that $f$ is a function will disallow (or will
`conflict' with) $f(r_0)= w_2$.  Such a morphism need not be total over
$\esevents_0$.  The events for which $f$ is defined are called the
\defw{synchronisation events}; thus, reads are synchronisation events. 
Recall that synchronisation events are to be seen as events that
\emph{consume} other (free) events.  The events for which $f$ is undefined
are called \defw{free events}.  Writes are free events as they can happen
freely without having to synchronise with events from another event
structure.  We do \emph{not} require these morphisms to be injective: this
allows for multiple reads to read from the same write.  We require such a
morphism to be consistency preserving: that is, morphisms map consistent
histories in $\esconf_{1}$ to consistent histories in $\esconf_{2}$.  We
require that the morphisms preserve the $\enables$ relation as well.

The next three requirements capture the idiosyncrasies of shared memory. 
First, we require that a morphism preserves labels.  The labels are
understood to be as addresses of program variables: this ensures
that read and write operations can synchronise if and only if they are
performed on the same address/label.  Second, we demand that a morphism
never reads a value that is not written: that is, any write that a read
reads must have happened before the read.  The final requirement ensures
that a read always reads the latest write.

\hfill\\\noindent\textsc{product $\times$}: 
\label{sec:tc:ges:ops}
Let $\Gamma_{0} =  \gestruct{\esevents_{0}, \esconf_{0}, \enables_{0}}$ and
$\Gamma_{1} = \gestruct{\esevents_{1}, \esconf_{1}, \enables_{1}}$ be two
stable event structures.  The \defw{product} $\Gamma = \gestruct{\esevents,
\esconf, \enables}$, denoted $\mathbf{ \Gamma_{0}\times\Gamma_{1}}$, is
defined as follows:

\begin{itemize}
\renewcommand{\labelitemi}{$-$}

\item $\esevents \define \set{(e_{0},*) \:\:|\: e_{0} \in \esevents_{0}}$
\phantom{1} $\bigcup$ \phantom{1} $\,\set{(*,e_{1}) \:|\: e_{1} \in
\esevents_{1}}$ \phantom{1} $\bigcup$ \\ \phantom{1} $\quad\,
\set{(e_{0},e_{1}) \:|\: e_{0} \in \esevents_{0}$,$ e_{1} \in
\esevents_{1}$, $\eslfn(e_{0}) = \eslfn(e_{1})}$

\item Let the \defw{projection morphisms} $\pi_{i}: \esevents \to
\esevents_{i}$ be defined as $\pi_{i}(e_{0}, e_{1}) = e_{i}$, for $i = 0,
1$.  Using these projection morphisms, let us now define the $\esconf$ of
the product event structure as follows: for $X \subseteq \esevents$,
we have $X \in \esconf$ when
\begin{itemize}
\item $\{X \:|\: X \subseteq_\mathit{finite} \esevents \}$  

\item  $\pi_{0}X \in \esconf_{0}$, $\pi_{1}X \in \esconf_{1}$

\item Read events in $X$ \defw{form a function}: $\forall e, e' \in X$, $\big( (\pi_{0}(e) = \pi_{0}(e') \ne *) \wedge (\pi_{0}(e) \in \esread_0 )\big) \vee \big(
(\pi_{1}(e) = \pi_{1}(e') \ne *) \wedge (\pi_{1}(e) \in \esread_1)\big)
 \Rightarrow$ $e = e'$

\item events in $X$ \defw{do not time travel}: that is, $\forall e \in X$, $\big((\pi_0(e) \in \eswrite_0 \wedge \pi_1(e) \in \esread_1) \Rightarrow  \clock{\pi_0(e)}{\pi_{0}X}\footnote{$\clock{w_i}{\pi_iX}$ denotes the clock value of the event $w_i$ in $\pi_iX$;  $i$ denotes the index of the process/thread that issued $w_i$. Note that our clock constraints only restrict the clocks of per-thread events, and the clock values of the product events are left `free'.} < \clock{\pi_1(e)}{\pi_{1}X}\big)$ $\:\wedge\:$
$\big((\pi_0(e) \in \esread_0 \wedge \pi_1(e) \in \eswrite_1) \Rightarrow  \clock{\pi_1(e)}{\pi_{1}X} < \clock{\pi_0(e)}{\pi_{0}X}\big)$

\item read events in $X$ \defw{read the latest write}: $\forall e \in X$,  
$\pi_0(e) \in \eswrite_0 \wedge \pi_1(e) \in \esread_1$, 

$\forall w_i \in  \eswrite_0(\eslfn(\pi_0(e)))\footnote{$\eswrite_0(\eslfn(\pi_0(e)))$ denotes the set of write events in thread $0$ that share the same address/label as $\pi_0(e)$.} \setminus \pi_0(e)$,

$\clock{\pi_1(e)}{\pi_1X} > \clock{w_i}{\pi_iX}$\\
\phantom{10ex}
$\Rightarrow \clock{\pi_0(e)}{\pi_0X} > \clock{w_i}{\pi_iX}$\footnote{The dual of this requirement, where we swap $0$ and $1$, is also assumed; we omit stating it for brevity.}

\item in any given $X$, all the write events to the same address are \defw{totally ordered}. Let  $\Sigma$ be a finite, label set, denoting the set of addresses/variables in the program. Then,

$\forall l \in \bigcup_i \Sigma_i$, $i\in \setof{0,1}$, $\forall w, w' \in \eswrite(l)$, $(\clock{w}{\pi_iX} < \clock{w'}{\pi_iX}) \vee (\clock{w'}{\pi_iX} < \clock{w}{\pi_iX})$

\end{itemize}
\item  $X \enables e \define \forall X \in \esconf$, $\forall e \in \esevents$, $ 0 \le i, j \le 1$, $i \ne j$, $e_i = \pi_i(e)$, $e_j = \pi_j(e)$,

$\big( e_i \in R_i \Rightarrow e_j \ne * \big)$ $\wedge$ %
$\big(e_i = * \wedge e_j \in W_j \Rightarrow \pi_{j} X \enables_{j} e_j \big)$ $\wedge$\\
$\big(e_i = * \wedge e_j \in IRW_j\footnote{The set $IRW_j \subseteq E_j$ denotes the set of internal/local reads in thread $j$: $IRW_j = \set{RW_{lm}\:|\:r_l \in R_j, w_m \in W_j, \eslfn(r_l) = \eslfn(w_m)}$.}\Rightarrow \pi_{j} X \enables_{j} e_j \big)$ $\wedge$\\
$\big(e_i \in R_i \wedge e_j \in W_j \Rightarrow \pi_{i} X \enables_{i} e_i  \wedge \pi_{j} X \enables_{j} e_j  \big)$\\

\end{itemize}

Products are a means to build larger event structures from components.  The
event set of product event structure has all the combinations of the
constituent per-thread events to account for all possible states of the
system.  A product should also accommodate the case where events in a thread
do not synchronise with any event in other threads.  This is ensured by
introducing the dummy event `$*$'.

We next demand that admissible executions in the product event structure
yield admissible executions in the constituent, per-thread event structures. 
This is ensured by introducing projection morphisms that `project' 
executions of the product event structure to their respective, per-thread
ones: we require these projected, per-thread executions to be consistent
executions.  Next, we forbid any read in an execution to match with more
than one write, ensure that a read's clock is greater than that of the
corresponding write's clock, and that a read always reads the latest
write.  We also demand that the writes to an address are always totally
ordered.  Finally, we demand that the enabling relation of product reflects
all the per-thread enabling relations.  This is ensured by requiring any
product-wise enabling yields a valid per-thread enabling.  It is important
to note that every event in the product event structure is a free event, and 
product events do not synchronise with any other event.

\vspace{-1ex}
\hfill\\\noindent\textsc{restriction $\lceil$}: 
Let $\Gamma = \gestruct{\esevents, \esconf, \enables}$ be an event structure. Let $A \subseteq \esevents$. We define the \defw{restriction} of $\Gamma$ to $A$, denoted $\mathbf{\esrestr{\Gamma}{A}} \define \gestruct{\esevents_{A}, \esconf_{A}, \enables_{A}}$, as follows.

\begin{itemize}
\renewcommand{\labelitemi}{$-$}
\item $\esevents_{A} \define A$
\item $X \in \esconf_{A} \Leftrightarrow X \subseteq A$, $X \in \esconf$
\item $X \enables_{A} e \define X \subseteq A$, $e \in A$, $X \enables e$

\end{itemize}

Restriction builds a new event structure containing only  events named in
the restriction set: it restricts the set of events to $A$, isolates
consistent sets involving events in $A$, and ensures that events of $A$ are
enabled appropriately.

%
\ifx 0
\cref{fig:wmm-bmc:eg:th1,fig:wmm-bmc:eg:th2}~of~\cref{fig:tc:eg} show a program fragment from~\cite{wmm:ver} that is adopted from PostgreSQL~\cite{pgsql}. The example shown has two worker threads that communicate using the four shared variables, {\tt latch1}, {\tt latch2}, {\tt flag1}, and {\tt flag2}. On the left is the corresponding intermediate form we have computed. The system has two per-thread \ppo{}s, defined over the read (with prefix {\tt R}) and write (with prefix {\tt W}) events; the subscripts indicate the SSA index. The directed arrows depict the intra-thread \ppo{}. The inter-thread lines depict the potential matches relation and colour of the line indicates the address/label of the program variable. The initial writes and the guards of the events are omitted for brevity. The bottom half of \cref{fig:tc:eg} shows the event set $\esevents$, and the enabling relation $\enables$ of the product event structure that correspond to the example. The product event structure is a fairly complex object and we only show an abridged version here: the event set omits the initial writes; a wildcard `?' serves as place-holder for a  suitable match; the `$\dots$' in the enablings denote the left-closed prefix of the event that succeeds it.

\begin{figure*}
\ganeshP{the worked out eventstructures go here}
\begin{scriptsize}
\begin{align*}
\esevents &= \{ 
\tuple{*, Wlatch1_6},
\tuple{Rlatch1_{2}, Wlatch1_6},
\tuple{Rlatch1_{3}, Wlatch1_6},  
\tuple{Rlatch1_{4}, Wlatch1_6}, 
\tuple{*, Wflag1_5}, 
\tuple{Rflag1_{2}, Wflag1_5}, 
\tuple{Rflag1_{3}, Wflag1_5},
\\&
\tuple{Wlatch2_2, *}, 
\tuple{Wlatch2_2, Rlatch2_3},
\tuple{Wlatch2_2, Rlatch2_{4}}, 
\tuple{Wlatch2_2, Rlatch2_{5}},
\tuple{Wflag2_2, *},
\tuple{Wflag2_2, Rflag2_{3}}, 
\tuple{Wflag2_2, Rflag2_{4}},
\\&
\tuple{*, Wflag2_5},
\tuple{*, Wlatch2_6}, 
\tuple{Wlatch1_5, *}, 
\tuple{Wflag1_4, *}
\}
\\
\enables &=
\set{\tuple{Rlatch1_{2}, ?}} \mathbf{\enables} \tuple{Rlatch1_{3}, ?}, 
\set{\dots, \tuple{Rlatch1_{3}, ?}} \enables \tuple{Rlatch1_{4}, ?}, 
\set{\dots, \tuple{Rlatch1_{4}, ?}} \enables \tuple{Rflag1_2, ?}, 
\set{\dots, \tuple{Rflag1_2, ?}} \enables \tuple{Wlatch1_5, *},
\\&
\set{\dots, \tuple{Rflag1_2, ?}} \enables \tuple{Rflag1_3, ?},
\set{\dots, \tuple{Wlatch1_5, *}, \tuple{Rflag1_3, ?}} \enables \tuple{Wflag1_4, *},
\set{\dots, \tuple{Wflag1_4, *}} \enables \tuple{Wflag2_2, *}, \dots\dots
\end{align*}
\end{scriptsize}
\caption{The event structures}
\label{fig:tc:eg}
\end{figure*}

\fi


\ifx 0
\lstset{language=C,numbers=none,  basicstyle=\ttfamily\scriptsize, float=t}
\begin{figure*}[t]
\centering
\begin{minipage}{0.495\linewidth}
\centering
\begin{minipage}{0.495\linewidth}
\begin{scriptsize}
\renewcommand{\lstlistingname}{PGSQL}
\begin{lstlisting}[caption=worker, label=fig:wmm-bmc:eg:th1]
for(int i=0;i<n;i++){
 for(int j=0;!latch1 && j<n;j++);
 if(!latch1) return NULL;
 assert(!latch1 || flag1);
 latch1 = false;
 if(flag1){
   flag1 = false;
   flag2 = true;
   //PSO_FENCE();
   latch2 = true;
 }
}
\end{lstlisting}
\end{scriptsize}
\end{minipage}
\begin{minipage}{.495\linewidth}
\begin{scriptsize}
\renewcommand{\lstlistingname}{PGSQL}
\begin{lstlisting}[caption=worker, label=fig:wmm-bmc:eg:th2]
for(int i=0;i<n;i++){
 for(int j=0;!latch2 && j<n;j++);
 if(!latch2) return NULL;
 assert(!latch2 || flag2);
 latch2 = false;
 if(flag2){
   flag2 = false;
   flag1 = true;
   //PSO_FENCE()
   latch1 = true;
 }
}
\end{lstlisting}
\end{scriptsize}
\end{minipage}
\end{minipage}
%
%
\begin{minipage}{0.45\linewidth}
\tikzset{
    punkt/.style={
           rectangle,
			fill=white,
 			blur shadow={shadow blur steps=3},
           draw=black, 
           text width=4em,
           minimum height=1ex,
           text centered},
    pil/.style={
           ->,
           shorten <=2pt,
           shorten >=2pt,}
}
\vspace{-3ex}
\begin{adjustbox}{scale=0.72}
\begin{scriptsize}
\begin{tikzpicture}[->,>=stealth',shorten >=1pt,auto,node distance=6.5em,
                    semithick,  remember picture]
  \tikzstyle{every state}=[fill=none,draw=black,text=black]

  \node[punkt]  (11)               {\tt Rlatch1$_2$};  
  \node[punkt]  (12) [right of=11] {\tt Rlatch1$_3$};
  \node[punkt]  (13) [right of=12] {\tt Rlatch1$_4$};
  \node[punkt]  (14) [right of=13] {\tt Rflag1$_2$};
  \node[punkt]  (15) [above right=2ex and 2ex of 14] {\tt Wlatch1$_5$};   
  \node[punkt]  (16) [below right=2ex and 2ex of 14] {\tt Rflag1$_3$};  
  \node[punkt]  (17) [below right=2ex and 2ex of 15] {\tt Wflag1$_4$}; 
  \node[punkt]  (18) [right of=17] {\tt Wflag2$_2$};
  \node[punkt]  (19) [right of=18] {\tt Wlatch2$_2$};
  
  \node[punkt]  (21) [below=21ex of 11] {\tt Rlatch2$_3$};
  \node[punkt]  (22) [right of=21] {\tt Rlatch2$_4$};
  \node[punkt]  (23) [right of=22] {\tt Rlatch2$_5$};
  \node[punkt]  (24) [right of=23] {\tt Rflag2$_3$};
  \node[punkt]  (25) [below right=2ex and 2ex of 24] {\tt Wlatch2$_6$};  
  \node[punkt]  (26) [above right=2ex and 2ex of 24] {\tt Rflag2$_4$};
  \node[punkt]  (27) [above right=2ex and 2ex of 25] {\tt Wflag2$_5$};
  \node[punkt]  (28) [right of=27] {\tt Wflag1$_5$};
  \node[punkt]  (29) [right of=28] {\tt Wlatch1$_6$};

  \path (11) edge    node {} (12)
        (12) edge    node {} (13)
        (13) edge    node {} (14)
        (14) edge    node {} (15)
        (14) edge    node {} (16)
        (15) edge    node {} (17)
        (16) edge    node {} (17)
        (17) edge    node {} (18)
        (18) edge    node {} (19)
        
        (21) edge    node {} (22)
        (22) edge    node {} (23)
        (23) edge    node {} (24)
        (24) edge    node {} (25)
        (24) edge    node {} (26)
        (25) edge    node {} (27)
        (26) edge    node {} (27)
        (27) edge    node {} (28)
        (28) edge    node {} (29);

        \draw[-, out=-45, in=-45,  red] (11) -- (29);
        \draw[-, red] (12) -- (29);
        \draw[-,  red] (13) -- (29);
        \draw[-,  red] (13) -- (29);
        \draw[-,  blue] (14) -- (28);
        \draw[-,  blue] (16) -- (28);
        \draw[-,  green] (21) -- (19);
        \draw[-,  green] (22) -- (19);
        \draw[-,  green] (23) -- (19);
        \draw[-,  orange] (24) -- (18); 
        \draw[-,  orange] (26) -- (18);
          
\end{tikzpicture}
\end{scriptsize}
\end{adjustbox}
\end{minipage}
\\
\vspace{1ex}
\begin{scriptsize}
\begin{align*}
\esevents &= \{ 
\tuple{*, Wlatch1_6},
\tuple{Rlatch1_{2}, Wlatch1_6},
\tuple{Rlatch1_{3}, Wlatch1_6},  
\tuple{Rlatch1_{4}, Wlatch1_6}, 
\tuple{*, Wflag1_5}, 
\tuple{Rflag1_{2}, Wflag1_5}, 
\tuple{Rflag1_{3}, Wflag1_5},
\\&
\tuple{Wlatch2_2, *}, 
\tuple{Wlatch2_2, Rlatch2_3},
\tuple{Wlatch2_2, Rlatch2_{4}}, 
\tuple{Wlatch2_2, Rlatch2_{5}},
\tuple{Wflag2_2, *},
\tuple{Wflag2_2, Rflag2_{3}}, 
\tuple{Wflag2_2, Rflag2_{4}},
\\&
\tuple{*, Wflag2_5},
\tuple{*, Wlatch2_6}, 
\tuple{Wlatch1_5, *}, 
\tuple{Wflag1_4, *}
\}
\\
\enables &=
\set{\tuple{Rlatch1_{2}, ?}} \mathbf{\enables} \tuple{Rlatch1_{3}, ?}, 
\set{\dots, \tuple{Rlatch1_{3}, ?}} \enables \tuple{Rlatch1_{4}, ?}, 
\set{\dots, \tuple{Rlatch1_{4}, ?}} \enables \tuple{Rflag1_2, ?}, 
\set{\dots, \tuple{Rflag1_2, ?}} \enables \tuple{Wlatch1_5, *},
\\&
\set{\dots, \tuple{Rflag1_2, ?}} \enables \tuple{Rflag1_3, ?},
\set{\dots, \tuple{Wlatch1_5, *}, \tuple{Rflag1_3, ?}} \enables \tuple{Wflag1_4, *},
\set{\dots, \tuple{Wflag1_4, *}} \enables \tuple{Wflag2_2, *}, \dots\dots
\end{align*}
\end{scriptsize}
\caption{An example program, its \tso Intermediate form and, the associated product event structure}
\label{fig:tc:eg}
\end{figure*}
\fi


\begin{lrbox}{\egestructbox}
\scriptsize
\centering
\noindent
\begin{minipage}{2\linewidth}
\tikzset{
    punkt/.style={
           rectangle,
			fill=white,
           draw=black,
           text width=3.4em,
           minimum height=0.25ex,
           text centered},
    pil/.style={
           shorten <=2pt,
           shorten >=2pt,}
}
\noindent
\flushleft
\begin{tikzpicture}	[trim left, ->,>=stealth',shorten >=1pt,auto,
					]
  \node[draw=none]	(0) at (0.5,3) 		   		{};

  \node[punkt]	(01)  [above=4em of 0]		   		{$\emptyset$ };
  \node[punkt]	(wx3) [above right=1em and 1em of 01] 	{$Wx_3$ };
  \node[punkt]	(ry3) [below right=1em and 1em of 01] 	{$Ry_3$ };
  \node[draw=none]	(001)  [right=7em of 01] 			{$\circ$ };

  \node[punkt]  (wx4) [above right=1em and 3em of 001]	{$Wx_4$};
  \node[punkt]  (rwx64) [right=3em of wx4] {$RWx_{64}$};

  \node[punkt]  (rx6) [right=10em of 001] {$Rx_6$};

  \node[punkt]  (wx5) [below right=1em and 3em of 001]  {$Wx_5$};
  \node[punkt]  (rwx65) [right=3em of wx5] {$RWx_{65}$};

  \node[draw=none]  (0001) [right=17em of 001] {$\circ$};
  \node[punkt] (wy4)  [right=2em of 0001]	{$Wy_4$};  

  \node[draw=none] (thread1) [above right =1em and -4em of wy4] {Thread 1};

  \node[draw=none] [above right=-3em and 1em of wy4] 
	{
    $\begin{aligned}
	\#_1	&=
			\set{(e_1, e_2) \:|\: e_1, e_2 \in \esevents_1	\wedge \pathcond(e_1) \wedge  \pathcond(e_2) \ne \bot}
			\:\:\bigcup\\
			&\quad\:\:
			\set{(RWx_{64}, RWx_{65}), (Rx_6, RWx_{64}), (Rx_6, RWx_{65})}\\
	\Sigma_1 &= \set{x, y}
      \end{aligned}$
	};
%
  \node[punkt]	(02) 	 [below=4em of 0]		   		{$\emptyset$ };
  \node[punkt]	(wy5) [above right=1em and 1em of 02] 	{$Wy_5$ };
  \node[punkt]	(rx7) [below right=1em and 1em of 02] 	{$Rx_7$ };
  \node[draw=none]	(002)  [right=7em of 02] 			{$\circ$ };

  \node[punkt]  (wy6) [above right=1em and 3em of 002]	{$Wy_6$};
  \node[punkt]  (rwy86) [right=3em of wy6] {$RWy_{86}$};

  \node[punkt]  (ry8) [right=10em of 002] {$Ry_8$};

  \node[punkt]  (wy7) [below right=1em and 3em of 002]  {$Wy_7$};
  \node[punkt]  (rwy87) [right=3em of wy7] {$RWy_{87}$};

  \node[draw=none]  (0002) [right=17em of 002] {$\circ$};
  \node[punkt] (wx8)  [right=2em of 0002]	{$Wx_8$};  

  \node[draw=none] (thread2) [above right =1em and -4em of wx8] {Thread 2};

  \node[draw=none] [above right=-3em and 1em  of wx8] 
	{
     $\begin{aligned}
	\#_2	&=	
				\set{(e_1, e_2) \:|\: e_1, e_2 \in \esevents_2	\wedge \pathcond(e_1) \wedge  \pathcond(e_2) \ne \bot}
				\:\:\bigcup\\
			&\quad\:\:
			\set{(RWy_{86}, RWy_{87}), (Ry_{8}, RWy_{86}), (Ry_{8}, RWy_{87})}\\
	\Sigma_2 &= \set{x, y}
      \end{aligned}$
	};
  \path 
		(01)  	edge    node {} (wx3)
  		(01)	edge    node {} (ry3)
  		(wx3)	edge    node {} (001)
  		(ry3)	edge    node {} (001)

  		(001)   edge [red, dashed]    node {} (wx4)
  		(wx4)  	edge [red, dashed]    node {} (rwx64)
 		(wx4)  	edge [red, dashed]    node {} (rx6)
		(rwx64) edge [red, dashed]    node {} (0001)

  		(001)   edge [red, dashed]    node {} (wx5)
  		(wx5)  	edge [red, dashed]    node {} (rwx65)
 		(wx5)  	edge [red, dashed]    node {} (rx6)
		(rwx65) edge [red, dashed]    node {} (0001)

		(rx6)  	edge [red, dashed]    node {} (0001)
		(0001)  edge  node {} (wy4)
		(02)	edge    node {} (wy5)
  		(02)  	edge    node {} (rx7)
  		(wy5)  	edge    node {} (002)
  		(rx7)  	edge    node {} (002)

  		(002)   edge [red, dashed]    node {} (wy6)
  		(wy6)	edge [red, dashed]    node {} (rwy86)
 		(wy6)	edge [red, dashed]    node {} (ry8)
		(rwy86)	edge [red, dashed]    node {} (0002)

  		(002)   edge [red, dashed]    node {} (wy7)
  		(wy7)	edge [red, dashed]    node {} (rwy87)
 		(wy7)	edge [red, dashed]    node {} (ry8)
		(rwy87)	edge [red, dashed]    node {} (0002)

		(ry8)	edge [red, dashed]    node {} (0002)
		(0002)  edge  node {} (wx8)
				;
\end{tikzpicture}
\end{minipage}%
\end{lrbox}

\begin{figure*}
\centering
\begin{subfigure}{\linewidth}
\centering
\usebox{\egestructbox}
\caption{The per-thread event structures}
\label{fig:sem:eg:per-thread}
\end{subfigure}
\\
\noindent
\begin{subfigure}{\linewidth}
\centering
\begin{align*}
\esevents &= \{
				(Ry_3, Wy_5),  (Ry_3, Wy_6), (Ry_3, Wy_7), (RWx_{64}, *), 
				(RWx_{65}, *), (Rx_6, Wx_8),\\
		  &\qquad 
				(Wx_3, Rx_7),  (Wx_4, Rx_7),  (Wx_5, Rx_7), (*, RWy_{86}), 
				(*, Ry_{87}), (Wy_4, Ry_8),
			\}
		\qquad\qquad\quad\:\:\:
		\Sigma = \Sigma_1 \cup \Sigma_2 = \set{x, y}\\
\#		  &= \{((R-_i, W-_j), (R-_i, W-_k)) \;|\; (R-_i, W-_j), (R-_i, W-_k) \in \esevents, - \in \Sigma\}
\end{align*}
\caption {Semantics of the shared memory program}
\label{fig:sem:eg:product}
\end{subfigure}
\caption{Event structure constructions for the example given in~\cref{fig:wmm-bmc:eg}}
\label{fig:sem:eg}
\end{figure*}

\section{Semantics for weak memory}
\label{sec:sem}

Let $P$ be a shared memory program with $n$ threads. Each thread is modelled
by an event structure $\Gamma_{i} = \gestruct{\esevents_{i}, \esconf_{i},
\enables_{i}}$, which we call a~\emph{per-thread event structure}.  The
per-thread event structures are constructed using our per-thread \ppo{}s and
the guards associated with the read/write events.  The computed guards
naturally carry the control and data flow choices of a thread into the
conflict relation of the corresponding per-thread event structure: two
events are conflicting if their guards are conflicting; conflicting guards
are those that cannot together be set to true.

As we build our $\esevents_i$ from $\ppo_i$, in addition to all the
read/write events in the $\ppo_i$, we also add the set of local reads
($\ireads _i$) (of thread $i$) into $E_i$ as free events; we call a read
event $RWx_{kl}$ a \emph{local} or \emph{internal read} if it reads from a
write $Wx_l$ from the same thread.  Note that all possible write events that
can feed a value to a given local read can be statically computed (e.g.,
using def-use chains).  Such local reads (say $RWx_{kl}$) are added as free
events in $E_i$: in doing so, we require that the guards of the constituent
events (\pathcond($RWx_{kl}$) and \pathcond($Wx_l$)) do not conflict, and
that functoriality/freshness of reads/writes is guaranteed in all per-thread
$X \in \esconf_i$ involving them.  The intuition is that reads reading from
local writes are free to do so without `synchronising' with any other
thread.  Our \potmat relation is constructed after adding such free, local
reads.  Since a read event can either read a local write or a (external)
write from another thread, local reads are considered to be in conflict with
the external reads that has to `synchronise' with other threads.  This
conflict captures the fact that at runtime only one of these events will
happen.

Let us also denote the system-wide \potmat as~$\potmatm$. We are now in a
position to define our \defw{truly concurrent semantics for the shared
memory program} $P$: the system of the $n$-threaded program $P$ is
over-approximated by $\semantics{P} \define \Gamma_{P} =
\gestruct{\esevents, \esconf, \enables} \define
\Big(\esrestr{\prod_{i=0}^{n-1} \gestruct{\esevents_{i}, \esconf_{i},
\enables_{i}}\Big)}{\potmatm}$.  This compositional, conflict-aware, truly
concurrent semantics for multi-threaded shared memory programs, written for
modern weak memory architectures is a novel contribution.  Our symbolic
product event structure faithfully captures the (abstract) semantics of the
multi-threaded shared memory program: since it is conflict-aware, this
semantics can also distinguish systems at least up to `failure
equivalence'~\cite{roscoe1998theory}, whereas coarser partial order based
semantics like~\cite{cav13} can only distinguish systems up to trace
equivalence.

\hfill\\\noindent\textsc{An Example}:
\cref{fig:sem:eg} depicts the event structure related constructs for the
example given in~\cref{fig:wmm-bmc:eg}.  The top row,
\cref{fig:sem:eg:per-thread}, gives the per-thread event structures.  The
nodes depict the events and the arrows depict the pre-thread enabling
relation; the dummy node `$\circ$' is added only to aid the presentation. 
The solid, black lines depict non-conflicting enablings while the dotted,
red lines show the enablings that are conflicting: for instance, in Thread
1, $\emptyset$ enables \emph{both} events $Wx_3$ and $Ry_3$\footnote{We are
omitting the corresponding internal read $RWy_{30}$ for brevity, which may
capture the potential read from initial write $Wy_0$, capturing a read from
an uninitialized value (Ref.~section~\ref{sec:pr}) on SSA.}, while
$\set{\emptyset, Wx_3, Ry_3}$ enables only one of $Wx_4$ or $Wx_6$.  Note
that the added local read events $RWx_{64}$ and $RWx_{65}$ are mutually
conflicting, and these local reads in turn conflict with the $Rx_6$ event
that could be satisfied externally.  The conflict relation for both the
threads is given on the right hand side of the diagrams; the symmetries are
assumed.  The label set is given by (base) names of the \ssa variables: that
is, $\Sigma = {x, y}$; the labeling function is a natural one taking the
\ssa variables to their base name, forgetting the indices.  The bottom row,
\cref{fig:sem:eg:product}, gives the event set and conflict relation for our
semantics.  That is, it gives $\Gamma_{P} = \gestruct{\esevents, \esconf,
\enables} \define \Big(\esrestr{\prod_{i=0}^{n-1} \gestruct{\esevents_{i},
\esconf_{i}, \enables_{i}}\Big)}{\potmatm}$.  Note that $\esevents$ has only
those product events that are present in $\potmatm$; we omit the write
events of \potmatm for brevity.  In this slightly modified, but equivalent,
presentation we included the functoriality condition of the product into the
conflict relation.  That is, for every variable (as given by the label set
$\Sigma$), we demand that any read event synchronising with some write event
conflicts with the same read event synchronising with any other write event. 
We conspicuously omit presenting $\esconf$ as it is an exponential object
(in number of events): the elements of $\esconf$, apart from being conflict
free, are required to read the latest write, and that the reads do not time
travel.  The enabling relation of the final event structure relates an
element~$C$ of $\esconf$ with an element $e$ of $\esevents$ if the
per-thread projections of $C$ themselves enable all the participating
per-thread events in~$e$.
\subsection{Reachability in Weak Memory Programs}
\label{sec:sem:assert}

Having defined the semantics of weak memory programs, we now proceed to show
how we exploit this semantics to reason about valid executions/reachability
of shared memory programs.  Let $P = \tupleof{P_{i}}$, $0 \le i < n$ is a
shared memory system with $n$ threads.  Let $\semantics{P} \define
\Gamma_{P} = \gestruct{\esevents, \esconf, \enables} =
\Big(\esrestr{\prod_{i=0}^{n-1} \gestruct{\esevents_{i}, \esconf_{i},
\enables_{i}}\Big)}{\potmatm}$ be an event structure.

Let $\Gamma_{\natnum} = \tupleof{\esevents_{\natnum}, \esconf_{\natnum},
\enables_{\natnum}}$ be an event structure over natural numbers:
$\esevents_{\natnum} \:\define\: \natnum^0$; $\esconf_{\natnum} \:\define\:
\{\emptyset,\{\emptyset, 0\}, \{\emptyset, 0, 1\}, \cdots \}$; $\enables
\:\define\: \forall i \in \natnum^0 .  \{\emptyset, \cdots, i-1\}
\enables_{\natnum} i$.  We call this event structure a \defw{clock
structure}.  We would like to exploit the linear ordering provided by the
clock structure to `linearise' events in the product event structure; this
linearisation correspond to an execution trace of the system.  Naturally, we
would like this linearisation to respect the original event enabling order. 
This requirement is captured using a partial synchronous morphism from
$\Gamma_{P}$ into $\Gamma_{\natnum}$.  Let $\tau: \Gamma_{P} \to
\Gamma_{\natnum}$ be such a partial synchronous morphism.  Intuitively, such
a $\tau$ yields a `linear' execution that honours $\enables$ and $\esconf$. 
In other words, every match event is mapped (linearised) to an integer clock
position in the clock structure.  Each such $\tau$ yields a valid execution
of $\Gamma_P$.

Given such a $\tau$, let us now define a per-thread $\tau_{i}:\esevents_{i}
\to \esevents_{\natnum}$ as follows: $ \forall e \in \esevents,
\tau_{i}(e_{i}) = \tau(e), \phantom{1}\text{where} \phantom{1} e_{i} =
\pi_{i}(e) \phantom{1} \text {if}\ \defined(\pi_{i}(e))$; $\tau_{i}(e_{i})$
is undefined otherwise.  By construction each such $\tau_{i}$ yields a valid
execution for the thread~$i$.  Any per-thread assertion in the original
program can then be reasoned using such $\tau_i${}s: that is, a thread $i$
violates an assertion iff we have a $\tau_i$ (from a $\tau$) in which that
assertion is, first reached, and then violated.  In general, any (finite)
reachability related question can be answered using our semantics.

We say that a product event structure \defw{violates an assertion} whenever
in (at least) one of its secured maximal configurations, (at least) one of
its per-thread event structure's (projected) maximal configuration includes
the negation of the said assertion in that (secured) maximal configuration. 
The following theorem formalises this.

\begin{theorem}
\label{theorem:sem:assert}
The input program $P$ has an assertion violation iff there exists a maximal
$C \in \esconfigs{\esevents}$ such that at least one of
$\frestr{\tau_{i}}{C} \in \esconfigs{\esevents_i}$, $0 \le i < n$ contains
the negation of an assertion, under the specified memory model.
\label{estruct:lemma:assert}
\end{theorem}
%
The proof of the first part of the theorem (sans the memory model) follows
directly from construction.  Next, we present a proof sketch that addresses
memory model specific aspects.  Here we focus on \tso; suitable
strengthening/weakening (as dictated by \ppo) will yield a proof for
\seqc/\pso.

A shared memory execution is said to be a valid \tso execution if it
satisfies the following (informally specified) axioms~\cite{sparc:manual:8,
formal:sc-tso-pso}.  An execution is \tso iff:
\begin{enumerate}
\item \coherence: Write serialisation is respected.
\item \atomicity: The atomic load-store blocks behave atomically w.r.t.~other stores.
\item \termination: All stores and atomic load-stores terminate.
\item \val: Loads always return the latest write.
\item \storeOrd: Inter-store program order is respected.
\item \loadOrd: Operations issued after a load always follow the said load in the memory order.
\end{enumerate}

Our semantics omits \termination. Recall that $\clk$ denotes the clock
order.  Intuitively, the clock order represents the memory order.  Also,
recall that an execution corresponds to a (secured) maximal configuration. 
We now refer to the product definition~(see~\cref{fig:sem:eg:product}), and
show how the maximal configuration construction more or less directly
corresponds to these axioms.

The \coherence requirement is a direct consequence of demanding that writes
(to the same memory location) are totally ordered with respect to each
other.  The \atomicity axiom is enforced by assigning each atomic load-store
block the same clock value: this is done by making the atomic block elements
as incomparable/equal under \ppo.  The \val requirement is taken as is, with
time travel restrictions in the definition.  The \storeOrd and \loadOrd
requirements are enforced by \ppo, and are captured by the enabling
relation.  This shows that any valid maximal configuration respects \tso. 
The converse holds as the product simply uses the Cartesian product of
participating per-thread event sets and prunes it to the \tso specification.

This completes our sketch of the proof for \tso. Strengthening
of \loadOrd and weakening of \storeOrd{} --- via \ppo{} --- yield \seqc
and \pso, respectively.
%

\ifx 0

We add a last, but very crucial constraint on this product-restriction
construction.  We require that a read always reads the \defw{latest write}
from among all its potential matches.  That is, for a given read $r$, and
for all $\tuple{\cdots, r,\cdots, w', \cdots}$ in \potmat, if a
$\tuple{\cdots, r,\cdots, w, \cdots}$ is in a secured maximal configuration,
then for any $w' < r$, we require $w' <\footnote{In this abuse of notation,
the $<$ order is the less than order on integers, as provided by the
securing.} w$.  The events of $\Gamma_{P}$ are called matches and are all
free events.  The spawns and their associated read sets are treated as
separate sets to simplify the definitions.  But we expect the \ppo and the
potential matches relations to be suitably augmented to include
spawn-related events.

\fi

\begin{table*}
\begin{align*}
\ext\qquad  \phantom{2} &\define
	\bigwedge_{m=(r,w) \in \potmat;e \in \esevents(r) \cap \esevents(w)}
  	\bigg(
    	X_{e}
    	\Leftrightarrow
		\Big(
			\latestw(r, w) \phantom{1} \wedge \phantom{1} \funct(r, e)  \phantom{1} \wedge \phantom{1} \big(\pathcond(r) \wedge \pathcond(w) \big)
		\Big)
	\bigg)
\\
\nstep \qquad  \phantom{2} &\define \qquad\quad\,\,\,\,
	\bigwedge_{p \in \esevents_i; p  \enables_i^i q;}
	\qquad\:\:\,
	\quad
	\Big(
		\isbefore(C_{p}, C_{q})
	\Big)
\\
\mtoclk \qquad  &\define 
	\bigwedge_{m=(r, w)\in \potmat;e \in \esevents(r) \cap \esevents(w)}
	\bigg(
		X_{e} 	\Rightarrow 
\phantom{1} \isbefore(C_{w}, C_{r}) \phantom{1} \wedge \phantom{1} \isequal(V_{[r]}, V_{[w]})
	\bigg)
\\
\latestw(r, w) \qquad &\define \qquad
	\bigwedge_{(r,w') \in\potmat ; w \ne w'}
	\qquad
		\Big(
			(\neg \isbefore(C_{r}, C_{w'}) \wedge \pathcond(w'))
			\Rightarrow
			\neg \isbefore(C_{w}, C_{w'}) 
		\Big)
\\
\funct(q, m)\qquad &\define  \qquad \,\,\,\, 
	\bigwedge_{q \in \rset_i; e \in \esevents(q)\setminus m} 
	\qquad\quad
	\neg X_{e} 
\end{align*}
\textbf{A note on notations}: Empty conjunctions are interpreted as `true' and empty disjunctions as `false'. We read `$;$' as \emph{such that}: that is, `$e \in \fset;e \in E_i$' should be read as `$e \in \fset$ \emph{such that} $e \in E_i$'. We use the shorthand $\tuple{r,w}$ for the unique $n$-tuple $\tuple{\cdots,r,\cdots,w,\cdots}$ in $\potmatm$.

\end{table*}

\section{Encoding}
\label{sec:enc}

Let $\semantics{P} \define \gestruct{\esevents, \esconf, \enables} =
\Big(\esrestr{\prod_{i=0}^{n-1} \gestruct{\esevents_{i}, \esconf_{i},
\enables_{i}}\Big)}{\potmatm}$ be an event structure.  We 
build a propositional formula~$\Phi$ that is satisfiable iff the event
structure (hence the input program) has an assertion violation; $\Phi$ is
unsatisfiable otherwise.

The formula~$\Phi$ will contain the following variables: A Boolean variable
$X_{e}$ (for every $e \in \esevents$), a set of bit-vector variables $V_x$
(for every program variable~$x$) and a set of clock variables $C_{e_{ij}}$
(for every $e_{ij} \in \esevents_i$).  Given a per-thread event structure
$\Gamma_i = \gestruct{\esevents_i, \esconf_i, \enables_i}$, the
(conflict-free) \defw{covering relation}~\cite{estruct:ccs} of $\Gamma_i$
and the $\ppo_i$ coincide: informally, given a event structure over
$\esevents$, an event $e_1$ is covered by another event $e_2$ if $e_1 \ne
e_2$ and no other event can be `inserted' between them in any configuration. 
Let us denote this covering relation of the event structure $\Gamma_i$
as~$\enables_i^i$.  Intuitively, $\enables_i^i$ captures the per-thread
\emph{immediate} enabling, aka $\ppo_i$.  Let $\pathcond(e)$ denote the
guard of the event $e$.
 
Let $\assert$ be the set of program assertions whose violation we would like
to find; these are encoded as constraints over the $V_x$ variables.  Each
element of $\assert$ can be seen as a set of reads, with a set of values
expected from those reads, where the guards of all these reads evaluate to
true.  Equipped with $\enables_i^i$, $\pathcond(e)$, three types of
variables ($X_e$/$V_e$/$C_{e}$), and a set of \assert{}s, we now proceed to
define the formula $\Phi$ as follows:
\begin{align*}
\Phi \phantom{1} &\define \phantom{1} \ssae \wedge \ext \wedge \nstep \wedge \mtoclk \wedge \unique \wedge \neg (\bigwedge_i\assert_i)
\end{align*}

\noindent The constituents of $\Phi$ are discussed as below.
\begin{enumerate}

\item {\textbf{ssa ($\ssae$): }} These constraints include the intra-thread
\ssa data/control flow constraints; we rely on our underlying, modified
bounded model checker to generate them.

\item {\textbf{extension ($\ext$): }} A match can happen as soon as its
\pathcond{}s are set to true, its reads are not reused, and the read has
read the \latestw write.  The \funct constraint ensures that once a read is
matched, it cannot be reused in any other match; that is, it makes $f:R \to
W$ a function.  Thus, the \ext formula uses the enabling and conflict relation
to define left-closed, consistent configurations.

\item {\textbf{successors ($\nstep$): }} We require that the clocks respect
the per-thread immediate enabling relation.  This is the first step in
ensuring that configurations are correctly enabled and secured.

\item {\textbf{match2clock ($\mtoclk$): }} A match forces the clock values of
a write to precede that of the read (for non-local reads).  This ensures that
any write that a read has read from has already happened.  A~match also
performs the necessary data-flow between the reads and writes involved: that
is, a read indeed picks up the value written by the write.  The constraint
$\mtoclk$, together with $\nstep$, ensures that the $\latestw$ has the
expected, system-wide meaning; they together also lift the per-thread
enablings and securings to system-wide enablings and securings.

\item {\textbf{uniqueness ($\unique$): }} We require that the clocks of
writes that write to the same location are distinct.  Since the clock
ordering is total, this trivially ensures write serialisation.

\item {\textbf{clock predicates ($\isbefore$ and $\isequal$): }}
\label{sec:enc:clkpred}
The constraints $\isbefore$ and $\isequal$ define the usual $<$ and $=$ over
the integers.  We use bit-vectors to represent bounded integers.

\end{enumerate}

Let $k \define |\esread|+|\eswrite|$, where $|\esread|$ and $|\eswrite|$
denote the total number of reads and writes in the system.  That is, $k$ is
the total number of shared read/write events in the input program.  The
\defw{worst case cost} of our encoding is exactly $\frac{1}{4}k^2 + k \cdot
\log k$ Boolean variables; this follows directly from the
observation that each entry in the potential-matches relation can be seen as
an edge in the bipartite graph with vertex set $E = \esread \cup \eswrite$. 
Maximising for the edges (matches) in such a bipartite graph yields the
$\frac{1}{4}k^2$ component.  The $k \log k$ arises from the fact
that we need $k$ bit-vectors, each with $\log k$ bits to model the
clock variables.

\subsection{Soundness and Completeness: $\mathbf{\Phi \leftrightharpoons \Gamma_{P}}$}

The following theorems establish the soundness and completeness of the
encoding with respect to the assertion checking characterisation introduced
in \cref{sec:sem:assert}.  Note that any imprecision in computing the
potential-matches relation will not yield any false positives in $\Phi$, as
long as the \ppo is exact.  This is so even if $\potmat$ is simply the
Cartesian product of reads and writes.

\begin{theorem}
{[}Completeness: ${\Gamma_P \Rightarrow \Phi}${]}\\
For every assertion-violating event structure $\Gamma_{P} =
\gestruct{\esevents, \esconf, \enables} = \Big(\esrestr{\prod_{i=0}^{n-1}
\gestruct{\esevents_{i}, \esconf_{i}, \enables_{i}}\Big)}{\potmatm}$, there
exists a satisfying assignment for $\Phi$ that yields the required assertion
violation.
\label{theorem:estruct:to:sat}
\end{theorem}

\begin{theorem}
{[}Soundness: ${ \Phi \Rightarrow \Gamma_P}${]}\\
For every satisfying assignment for $\Phi$, there is a corresponding
assertion violation in the event structure $\Gamma_{P} =
\gestruct{\esevents, \esconf, \enables} = \Big(\esrestr{\prod_{i=0}^{n-1}
\gestruct{\esevents_{i}, \esconf_{i}, \enables_{i}}\Big)}{\potmatm}$.
\label{theorem:sat:to:estruct}
\end{theorem}

We omit the proofs owing to lack of space. Instead, we here provide the
intuition behind our encoding.  It is easy to see that $\ext$ and $\nstep$
together `compute' secured, consistent, per-thread configurations whose
initial events are enabled by the empty set: \funct guarantees every read
reads exactly one write and \latestw ensures that reads always pick up the
latest write (as ordered by \nstep) locally.  $\assert$ picks out all
secured, consistent configurations that violates any of the assertions.  But
an assignment satisfying these three constraints needs not satisfy the
system-wide \emph{latest write} requirement.  This is addressed by the
\mtoclk constraint: this constraint `lifts' the per-thread orderings to a
valid inter-thread total ordering; it also performs the necessary value
assignments.  Equipped with \mtoclk, the \latestw now can pick the writes
that are indeed latest across the system.  The transitivity of \isbefore (in
\nstep, \latestw, \mtoclk) correctly `completes' the per-thread orderings to
an arbitrary number of threads, involving arbitrary configurations of
arbitrary matches.

\ifx
\begin{proof}
Let $\eta=\ldots,m_{j},\ldots$; $m \in \esevents$, $0 \le j < p$ is a
non-zero deadlocking trace of length $p$ in $P$.  Let us call the each $m
\in \esevents$ a match.  By Theorem~\ref{estruct:lemma:dlock}, $\Gamma_{P}$
has a maximal configuration $C$ such that at least one of
$\frestr{\tau_{i}}{C}$, $0 \le i < n$ is a partially synchronous morphism. 
We derive the following satisfying assignment using such a $C$ and $\tau$.

We set $X_{m}$ to true if and only if $m \in C$. The configuration $C$ will
include at least one $m_{0}$ such that $\emptyset \enables m_{0}$ because
$C$ is a configuration and has a securing starting from $\emptyset$. 
Setting $X_{m_{0}}$ to true for all such $m_{0}$'s will satisfy $\bstrap$. 
$\ext$ is collectively satisfied as follows: conflict-freeness of $C$
ensures that only non-conflicting, per-process predecessor matches are
included in $C$; the consistency predicate of the product construction
ensures $\excl$ is satisfied; maximality of $C$ ensures that for any $m \in
C$, $C$ has all the events that are enabled by $m$ and are non-conflicting
with any events in $C$; thus $C$ satisfies $\ext$.
 
Since at least one of $\frestr{\tau_{i}}{C}$, is a partially synchronous
morphism, $\assert$ is satisfied.  We assign $\tau(e) = j$ for each $m_{j}$
in $\eta$: $\frestr{\tau_{i}}{C}$, $ 0 \le i < n$ thus satisfying $\mtoclk$
and $\smonotone$.
\end{proof}
\fi

\ifx
\begin{proof}
Notice that there is a one-to-one correspondence between the Boolean
variables involved in parts ($\ext$, $\dlock$) of $\Phi$ and the events of
$\Gamma_{P}$.  Let $C \subseteq \esevents$ be the set of such events that
are set to true in the given satisfying assignment.  We claim that $C$ is
configuration of $\Gamma_{P}$.  For every $m \in C$, $\excl$ ensures that
all the matches that conflict with $m$ are unset.  $\ext$ ensures that for
all $m \in C$, the immediate, non-conflicting predecessors (that enable $m$)
are also in $C$.  Also, for any event $m \in C$, following this chain of
immediate predecessors until $X_{m_{0}}$, where $X_{m_{0}}$ is part of
$\bstrap$ will give us a securing: $\emptyset \enables \{m_{0}\}, \ldots, X
\subset C \wedge X \in \esconf \enables m_{j}, \ldots, Y \in C \wedge X
\subseteq Y \wedge Y \in \esconf \enables m$.  Thus, $C$ is a configuration
of $\Gamma_{P}$.

We now claim that $C$ is a maximal configuration. Note that for every
$X_{m}$ that is set, $\ext$ forces all the events enabled by $m$ to be set
as well -- as long they do not conflict with any other events in $C$.  Thus,
$C$ has all the consistent set of events that could possibly occur together,
respecting $\esconf$.  Thus $C$ is a maximal configuration.

Given the set of clock variables $Clk_{\pi_{i}(m)}$ from the satisfying
assignment, we now define a $\tau: \Gamma_{P} \to \Gamma_{\natnum}$ as
follows.  $\tau(m) = Clk_{\pi_{i}(m)}$, if $def(\pi_{i}(m)) \wedge m \in C$
; undefined otherwise.  Now, $\mtoclk$ ensures that $\tau(m)$ is defined for
all $m \in C$.  Notice also that the particular $i$ does not matter as any
satisfying assignment to $\mtoclk$ requires that all the $Clk_{\pi_{i}(m)}$
be equal.  We claim that $\tau$ is a partially synchronous morphism. 
$\smonotone$ and $\mtoclk$ together ensure that $\tau$ preserves consistency
and enabling.  Thus $\tau: \Gamma_{P} \to \Gamma_{\natnum}$ is a partial
synchronous morphism.

Now, we claim that there is at least one $\frestr{\tau_{i}}{C}$ which is a
partial synchronous morphism from $\Gamma_{P}$ to $\Gamma_{\natnum}$.  This
is ensured by $\dlock$: for any $X_{f}$ (from, say process $i$) that is not
set $\frestr{\tau_{i}}{C}$ will be a partial synchronous morphism.

Thus, any satisfying assignment to $\Phi$ will result in a maximal
configuration $ C \in C_{max}$ such that at least one of
$\frestr{\tau_{i}}{C}$, $0 \le i < n$ is a partially synchronous morphism. 
By Theorem~\ref{estruct:lemma:dlock}, $P$ has a deadlocking trace given by
$\tau$.
\end{proof}
\fi

\section{Evaluation}
\label{sec:eval}

We have implemented our approach in a tool named~\tc\footnote{A hark back to
the conflict-friendly Norse decision makers; also an anagram of {\bf W}eak
memory {\bf A}na{\bf L}ysis using {\bf C}onfl{\bf I}ct aware t{\bf R}u{\bf
E} concurrenc{\bf Y}.} using \cbmc~\cite{cbmc} version~5.0 as a base.  The
tool currently supports the \seqc, \tso and \pso memory models.  \tc takes a
C/C++ program and a loop bound~$k$ as input, and transforms the input
program into a loop-free program where every loop is unrolled at least $k$
times.  From this transformed input program we extract the \ssa and the \ppo
and \potmat relations.  Using these relations, and the implicit event
structure they constitute, we build the propositional representation of the
input program.  This propositional formula is then fed to the MiniSAT
back-end to determine whether the input program violates any of the
assertions specified in the program.  If a satisfying assignment is found,
we report assertion violation and present the violating trace; otherwise, we
certify the program to be bug free, up to the specified loop bound~$k$.

We use two large, well established, widely used benchmark suites to evaluate
the efficacy of \tc: the Litmus tests from~\cite{litmus} and
SV-COMP~2015~\cite{sv-comp15} benchmarks.  We compare our work against the
state of the art tool to verify real-world weak memory programs,
\cite{cav13}; hereafter we refer it as \cav.  We remark that (\cite{cav13},
page 3) employs the term ``event structures'' to mean the per-processor
total order of events, as dictated by \textsc{po}.  This usage is unrelated
to \emph{our} event structures; footnote \#4 of~\cite{cav13} clarifies this
point.  We run all our tests with \emph{six} as the unrolling bound and
$900\,\mbox{s}$ as the timeout.  Our experiments were conducted on a 64-bit
3.07\,GHz Intel Xeon machine with 48\,GB of memory running Linux.  Out tool
is available at \url{https://github.com/gan237/walcyrie}; the URL provides
all the sources, benchmarks and automation scripts needed to
reproduce our results.

Litmus tests~\cite{litmus} are small~(60 LOC) programs written in a toy
shared memory language.  These tests capture an extensive range of subtle
behaviours that result from non-intuitive weak memory interactions.  We
translated these tests into C and used \tc to verify the resulting C code. 
The Litmus suite contains $5804$ tests and we were able to correctly analyse
all of them in under $5\,\mbox{s}$.

The SV-COMP's concurrency suite~\cite{sv-comp15} contains a range of
weak-memory programs that exercise many aspects of different memory models,
via the pthread library.  These include crafted benchmarks as well as
benchmarks that are derived from real-world programs: including device
drivers, and code fragments from Linux, Solaris, NetBSD and FreeBSD.  The
benchmarks include non-trivial features of C such as bitwise operations,
variable/function pointers, dynamic memory allocation, structures and
unions.  The \seqc/\tso/\pso part of the suite has 600 programs; please
refer to~\cite{sv-comp15} for the details.  \tc found a handful of
misclassifications (that is, programs with defects that are classified as
defect-free) among the benchmarks; there were no misclassifications in the
other direction, that is, all programs classified by the developers as
defective are indeed defective.  Such misclassifications are a strong
indication that spotting untoward weak memory interactions is tricky even
for experts.  We have reported these misclassifications to the SV-COMP
organisers.

The work that is closest to us is \cav~\cite{cav13}: like us, they use BMC
based symbolic execution to find assertion violations in C programs.  The
key insight here is that executions in weak memory systems can be seen as
partial orders (where pair of events relaxed are incomparable).  Based on
this, they developed a partial order based decision procedure.  Like us,
they rely on SAT solvers to find the program defects.  But our semantics is
conflict-aware, consequently the resulting decision procedure is also
different; the formula generation complexity for both approaches is cubic
and both of us generate quadratic number of Boolean variables.  The original
implementation provided with~\cite{cav13} handled thread creation
incorrectly.  We modified \cav to fix this, and we use this corrected
version in all our experiments.  Though the worst case complexity of both
approaches is the same, our true concurrency based encoding is more compact:
\tc often produced nearly half the number of Boolean variables and about 5\%
fewer constraints compared to \cav~(after 3NF reduction).


\begin{figure*}
\centering
\begin{subfigure}{0.33\textwidth}
\begin{tikzpicture}[domain=1e-2:1e4, scale=0.4]
\centering
\begin{loglogaxis}[%
	width=5.5in, 
    height=2.34in,
	xlabel=\tc (secs),
	ylabel=\cav (secs),
	compat = 1.3,
	ylabel shift = -10 pt,
	xlabel shift = -3 pt,
	ymin = .01e1, ymax=1e1,
	xmin = .01e1, xmax=1e1,
	xtick={.01e1, .1e1, 1e1},
	ytick={.01e1, .1e1, 1e1},
	scatter/classes={%
		unknown={mark=triangle*,red},%
		invalid={mark=triangle*,red},%
		valid={mark=x,blue}%
		}
	]

\addplot[smooth, thick] {x}; 
\addplot[scatter,only marks,%
		scatter src=explicit symbolic]%
		table [col sep=comma, x=TCusertime, y=CAVusertime, comment chars={\#},  meta=TCexpected] {plots/tc.results.cbmc.sc.csv-cav13.results.cbmc.sc.csv.both}; 
\end{loglogaxis}
\end{tikzpicture}
\captionof{figure}{Litmus:~\seqc}
\label{fig:eval:litmus:tc-cav13:sc:rtime}
\end{subfigure}
\begin{subfigure}{0.33\textwidth}
\begin{tikzpicture}[domain=1e-2:1e4, scale=0.4]
\centering
\begin{loglogaxis}[%
	width=5.5in,
    height=2.34in,
	xlabel=\tc (secs),
	ylabel=\cav (secs),
	compat = 1.3,
	ylabel shift = -10 pt,
	xlabel shift = -3 pt,
	ymin = .01e1, ymax=1e1,
	xmin = .01e1, xmax=1e1,
	xtick={.01e1, .1e1, 1e1},
	ytick={.01e1, .1e1, 1e1},
	scatter/classes={%
		unknown={mark=triangle*,red},%
		invalid={mark=triangle*,red},%
		valid={mark=x,blue}%
		}
	]

\addplot[smooth, thick] {x}; 
\addplot[scatter,only marks,%
		scatter src=explicit symbolic]%
		table [col sep=comma, x=TCusertime, y=CAVusertime, comment chars={\#},  meta=TCexpected] {plots/tc.results.cbmc.tso.csv-cav13.results.cbmc.tso.csv.both}; 
\end{loglogaxis}
\end{tikzpicture}
\captionof{figure}{Litmus:~\tso}
\label{fig:eval:litmus:tc-cav13:tso:rtime}
\end{subfigure}
\begin{subfigure}{0.33\textwidth}
\begin{tikzpicture}[domain=1e-2:1e4, scale=0.4]
\centering
\begin{loglogaxis}[%
	width=5.5in,
    height=2.34in,
	xlabel=\tc (secs),
	ylabel=\cav (secs),
	compat = 1.3,
	ylabel shift = -10 pt,
	xlabel shift = -3 pt,
	ymin = .01e1, ymax=1e1,
	xmin = .01e1, xmax=1e1,
	xtick={.01e1, .1e1, 1e1},
	ytick={.01e1, .1e1, 1e1},
	scatter/classes={%
		unknown={mark=triangle*,red},%
		invalid={mark=triangle*,red},%
		valid={mark=x,blue}%
		}
	]

\addplot[smooth, thick] {x};
\addplot[scatter,only marks,%
		scatter src=explicit symbolic]%
		table [col sep=comma, x=TCusertime, y=CAVusertime, comment chars={\#},  meta=TCexpected] {plots/tc.results.cbmc.pso.csv-cav13.results.cbmc.pso.csv.both}; 
\end{loglogaxis}
\end{tikzpicture}
\captionof{figure}{Litmus:~\pso}
\label{fig:eval:litmus:tc-cav13:pso:rtime}
\end{subfigure}
\\
\centering
\hspace{-0.5em}
\begin{subfigure}{0.33\textwidth}
\begin{tikzpicture}[domain=1e-2:1e4, scale=0.7]
\centering
\begin{loglogaxis}[%
	xlabel=\tc (secs),
	ylabel=\cav (secs),
	compat = 1.3,	
	ylabel shift = -5 pt,
	xlabel shift = -3 pt,
	ymin = 1e1, ymax=1.5e3,
	xmin = 1e1, xmax=1.5e3,
	xtick={1e1, 1e2, 1e3},
	ytick={1e1, 1e2, 1e3},
	scatter/classes={%
		SAT={mark=triangle*,red},%
		UNSAT={mark=x,blue}%
		}
	]
	
\addplot[smooth, thick] {x}; 
\addplot[dashdotted,thin] {900};
\node at (axis cs:1.75e1,900) [above] {900sec};
\draw [dashdotted, thin]  ({axis cs:900,0}|-{rel axis cs:0,0}) -- ({axis cs:900,0}|-{rel axis cs:0,1});
\addplot[scatter,only marks,%
		scatter src=explicit symbolic]%
		table [x=TCRT, y=CAVRT, comment chars={\#},  meta=EXPECTED] {plots/log.svcomp15.6.900s.cbmc.tc.reduct.minisat2.2.sc-log.svcomp15.6.900s.cbmc.cav13.trans.minisat2.2.sc.both}; 
\end{loglogaxis}
\end{tikzpicture}
\captionof{figure}{SV-COMP 2015: \seqc}
\label{fig:eval:svcomp:tc-cav13:sc:rtime}
\end{subfigure}
\begin{subfigure}{0.33\textwidth}
\begin{tikzpicture}[domain=1e-2:1e4, scale=0.7]
\centering
\begin{loglogaxis}[%
	xlabel=\tc (secs),
	ylabel=\cav (secs),
	compat = 1.3,
	ylabel shift = -5 pt,
	xlabel shift = -3 pt,
	ymin = 1e1, ymax=1.5e3,
	xmin = 1e1, xmax=1.5e3,
	xtick={1e1, 1e2, 1e3},
	ytick={1e1, 1e2, 1e3},
	scatter/classes={%
		SAT={mark=triangle*,red},%
		UNSAT={mark=x,blue}%
		}
	]

\addplot[smooth, thick] {x}; 
\addplot[dashdotted,thin] {900};
\node at (axis cs:1.75e1,900) [above] {900sec};
\draw [dashdotted, thin]  ({axis cs:900,0}|-{rel axis cs:0,0}) -- ({axis cs:900,0}|-{rel axis cs:0,1});
\addplot[scatter,only marks,%
		scatter src=explicit symbolic]%
		table [x=TCRT, y=CAVRT, comment chars={\#},  meta=EXPECTED] {plots/log.svcomp15.6.900s.cbmc.tc.reduct.minisat2.2.tso-log.svcomp15.6.900s.cbmc.cav13.trans.minisat2.2.tso.both}; 
\end{loglogaxis}
\end{tikzpicture}
\captionof{figure}{SV-COMP 2015: \tso}
\label{fig:eval:svcomp:tc-cav13:tso:rtime}
\end{subfigure}
\begin{subfigure}{0.33\textwidth}
\begin{tikzpicture}[domain=1e-2:1e4, scale=0.7]
\centering
\begin{loglogaxis}[%
	xlabel=\tc (secs),
	ylabel=\cav (secs),
	compat = 1.3,	
	legend entries={SAT, UNSAT},
	legend style={at={(.85, .21)}},	
	ylabel shift = -5 pt,
	xlabel shift = -3 pt,
	ymin = 1e1, ymax=1.5e3,
	xmin = 1e1, xmax=1.5e3,
	xtick={1e1, 1e2, 1e3},
	ytick={1e1, 1e2, 1e3},
	scatter/classes={%
		SAT={mark=triangle*,red},%
		UNSAT={mark=x,blue}%
		}
	]
	\addlegendimage{only marks, mark=triangle*,red}
	\addlegendimage{only marks, mark=x,blue}
	
\addplot[smooth, thick] {x}; 
\addplot[dashdotted,thin] {900};
\node at (axis cs:1.75e1,900) [above] {900sec};
\draw [dashdotted, thin]  ({axis cs:900,0}|-{rel axis cs:0,0}) -- ({axis cs:900,0}|-{rel axis cs:0,1});
\addplot[scatter,only marks,%
		scatter src=explicit symbolic]%
		table [x=TCRT, y=CAVRT, comment chars={\#},  meta=EXPECTED] {plots/log.svcomp15.6.900s.cbmc.tc.reduct.minisat2.2.pso-log.svcomp15.6.900s.cbmc.cav13.trans.minisat2.2.pso.both}; 
\end{loglogaxis}
\end{tikzpicture}
\captionof{figure}{SV-COMP 2015: \pso}
\label{fig:eval:svcomp:tc-cav13:pso:rtime}
\end{subfigure}
\\
\begin{subfigure}{0.33\textwidth}
\centering
\begin{tikzpicture}[domain=100:4000, scale=0.4]
\centering
\begin{loglogaxis}[%
	width=5.5in,
    height=2.34in,
	xlabel=\tc,
	ylabel=\cav,
	compat = 1.3,
	ylabel shift = -5 pt,
	xlabel shift = -3 pt,
	ymin = 100, ymax=4100,
	xmin = 100, xmax=4100,
	scatter/classes={%
		SAT={mark=triangle*,red},%
		UNSAT={mark=x,blue}%
		},
	log basis x=2,
	log basis y=2	
	]

\addplot[smooth, thick] {x}; 
\addplot[scatter,only marks,%
		scatter src=explicit symbolic]%
		table [
			x expr={\thisrow{TCpropagations}/\thisrow{TCconflicts}},
			y expr={\thisrow{CAVpropagations}/\thisrow{CAVconflicts}},
			comment chars={\#}, 
			meta=EXPECTED
		]
		{plots/log.svcomp15.6.900s.cbmc.tc.reduct.minisat2.2.sc-log.svcomp15.6.900s.cbmc.cav13.trans.minisat2.2.sc.table.both};
\end{loglogaxis}
\end{tikzpicture}
\captionof{figure}{\seqc: exploration efficacy~($\rho$)}
\label{fig:eval:tc-cav13:effic:sc}
\end{subfigure}
\begin{subfigure}{0.33\textwidth}
\centering
\begin{tikzpicture}[domain=100:4100, scale=0.4]
\centering
\begin{loglogaxis}[%
	width=5.5in,
    height=2.34in,
	xlabel=\tc,
	ylabel=\cav,
	compat = 1.3,
	ylabel shift = -5 pt,
	xlabel shift = -3 pt,
	ymin = 500, ymax=4100,
	xmin = 500, xmax=4100,
	scatter/classes={%
		SAT={mark=triangle*,red},%
		UNSAT={mark=x,blue}%
		},
	log basis x=2,
	log basis y=2
	]

\addplot[smooth, thick] {x}; 
\addplot[scatter,only marks,%
		scatter src=explicit symbolic]%
		table [
			x expr={\thisrow{TCpropagations}/\thisrow{TCconflicts}},
			y expr={\thisrow{CAVpropagations}/\thisrow{CAVconflicts}},
			comment chars={\#}, 
			meta=EXPECTED
		]
		{plots/log.svcomp15.6.900s.cbmc.tc.reduct.minisat2.2.tso-log.svcomp15.6.900s.cbmc.cav13.trans.minisat2.2.tso.table.both};
\end{loglogaxis}
\end{tikzpicture}
\captionof{figure}{\tso: exploration efficacy~($\rho$)}
\label{fig:eval:tc-cav13:effic:tso}
\end{subfigure}
\begin{subfigure}{0.33\textwidth}
\centering
\begin{tikzpicture}[domain=100:4000, scale=0.4]
\centering
\begin{loglogaxis}[%
	width=5.5in,
    height=2.34in,
	xlabel=\tc,
	ylabel=\cav,
	compat = 1.3,
	ylabel shift = -5 pt,
	xlabel shift = -3 pt,
	ymin = 500, ymax=4100,
	xmin = 500, xmax=4100,
	scatter/classes={%
		SAT={mark=triangle*,red},%
		UNSAT={mark=x,blue}%
		},
	log basis x=2,
	log basis y=2
	]

\addplot[smooth, thick] {x}; 
\addplot[scatter,only marks,%
		scatter src=explicit symbolic]%
		table [
			x expr={\thisrow{TCpropagations}/\thisrow{TCconflicts}},
			y expr={\thisrow{CAVpropagations}/\thisrow{CAVconflicts}},
			comment chars={\#}, 
			meta=EXPECTED
		]
		{plots/log.svcomp15.6.900s.cbmc.tc.reduct.minisat2.2.pso-log.svcomp15.6.900s.cbmc.cav13.trans.minisat2.2.pso.table.both};
\end{loglogaxis}
\end{tikzpicture}
\captionof{figure}{\pso: exploration efficacy~($\rho$)}
\label{fig:eval:tc-cav13:effic:pso}
\end{subfigure}
\caption{Evaluating \tc against \cav}
\end{figure*}


Our results are presented as scatter plots, comparing the total execution
times of \tc and \cav: this includes parsing, constraint generation and SAT
solving times; the smaller the time, the better the approach.  The $x$ axis
depicts the execution times for \tc and the $y$ axis depicts the same for
\cav.  The SAT instances are marked by a triangle
($\begingroup\color{red}\blacktriangle\endgroup$) and the UNSAT instances
are marked by a cross ($\begingroup\color{blue}\times\endgroup$).  The first
set of plots~(\cref{fig:eval:litmus:tc-cav13:sc:rtime},
\cref{fig:eval:litmus:tc-cav13:tso:rtime} and
\cref{fig:eval:litmus:tc-cav13:pso:rtime}) presents the data for the Litmus
tests.  There are three plots and each corresponds to the memory model
against which we tested the benchmarks.  Both \tc and \cav solve these small
problem instances fairly quickly, typically in under $5\,\mbox{s}$; recall
that individual Litmus tests are made of only tens of LOC.  But \cav appears
to have a slight advantage: we investigated this and found that \cav{}'s
formula generation was quicker while the actual SAT solving times were
comparable.  \tc{}'s current implementation has significant scope for
improvement: for instance, the \latestw and \funct constraint generation
could be integrated into one loop; also, the match generation could be
merged with the constraint generation.

The scatter plot \cref{fig:eval:svcomp:tc-cav13:sc:rtime} compares the
runtimes of \seqc benchmarks.  Under \seqc, the performance of \cav and \tc
appears to be mixed: there were seven (out of 125) UNSAT instances where \tc
times out while \cav completes the analysis between $10$ and
$800\,\mbox{s}$.  In the majority of the cases, the performance is
comparable and the measurements cluster around the diagonal.  Note that
\emph{no} modern multiprocessor offers \seqc, and the \seqc results are
presented for sake of completeness.

\cref{fig:eval:svcomp:tc-cav13:tso:rtime} presents the results of SV-COMP's
\tso benchmarks.  The situation here is markedly different, compared to the
Litmus tests and \seqc.  Here, \tc clearly outperforms \cav, as indicated by
the densely populated upper triangle.  This is contrary to usual intuition:
\tso, being weaker than \seqc, usually yields a larger state space, and the
conventional wisdom is that the larger the state space, the slower the
analysis.  Our results appear to contradict this intuition.  The same trend
is observed in the \pso section (\cref{fig:eval:svcomp:tc-cav13:pso:rtime})
of the suite as well.  In fact, \tc outperforms \cav over \emph{all} inputs. 

We investigated this further by looking deeper into the inner-workings of
the SAT solver.  SAT solvers are complex engineering artefacts and a full
description of their internals is beyond the scope of this article;
interested readers could consult~\cite{sat:handbook, sat:anatomy}.  Briefly,
SAT solvers explore the state space by making decisions (on the truth value
of the Boolean variables) and then propagating these decisions to other
variables to cut down the search space.  If a propagation results in a
conflict, the solver backtracks and explores a different decision.  There is a
direct  trade-off between propagations and conflicts, and a good encoding
balances these concerns judicially.  To this end, we introduce a metric
called \defw{exploration efficacy}, $\mathit{\rho}$, defined as the ratio
between the total number of propagations ($\mathit{prop}$) to the total
number of conflicts ($\mathit{conf}$).  That is, $\mathit{\rho} \define
\mathit{prop}/\mathit{conf}$.  The numbers $\mathit{prop}$ and
$\mathit{conf}$ are gathered only for those benchmarks where both the tools
provided a definite SAT/UNSAT answer.  Intuitively, one would expect SAT
instances to have a higher~$\rho$, while UNSAT instances are expected to
have a lower $\rho$.  To find a satisfying assignment, one needs to
propagate effectively (without much backtracking) and to prove
unsatisfiability, one should run into as many conflicts as early as
possible.  Thus, for SAT instances, a higher~$\rho$ is indicative of an
effective encoding; the converse holds true for UNSAT instances.

\cref{fig:eval:tc-cav13:effic:sc,fig:eval:tc-cav13:effic:tso,fig:eval:tc-cav13:effic:pso}
present the scatter plots for $\rho$ for \cav and \tc, for three of our
memory models.  For \seqc, the $\rho$ values are basically the same.  This
explains why we observed a very similar performance under \seqc.  The
situation changes with \tso and \pso.  The clustering of $\rho$ values on
the either side of the diagonal hints at the reason behind the superior
performance of our conflict-aware encoding.  In both \tso and \pso, our
$\rho$ values for the SAT instances are two to four times higher; our $\rho$
values for the UNSAT instances are one to two times lower.  In \pso, the
$\mathit{prop}$ values increase (as the state space grows with weakening)
and the number of conflicts $\mathit{conf}$ also grow in tandem, unlike
$\cav$.  We conjecture that this is the reason for the performance gain as
we move from \tso to \pso using \tc.

At the encoding level, the $\rho$ values can be explained by the way \tc
exploits the control and data-flow conflicts in the program.  Since \cav is
based on partial orders (which lack the notion of conflict), their encoding
relies heavily on SAT solver \emph{eventually} detecting a conflict.  That
is, \cav resolves the control and data conflicts \emph{lazily}.  By
contrast, \tc exploits the conflict-awareness of general event structures to
develop an encoding that handles conflicts \emph{eagerly}: branching time
objects like event structures are able to tightly integrate the control/data
choices, resulting in faster conflict detection and faster state space
pruning.  For instance, our $\funct$ constraint (stemming from the the
requirement that morphisms be functions) ensures that once a read ($r$) is
satisfied by a write $w$ (that is, when $X_{rw}$ is set to true; equally,
$f(r)=w$), all other matches involving the $r$ (say, $X_{rw'}$) are
invalidated immediately (via unit propagation).  This, along with the
equality in $\ext$, ensures that any conflicts resulting from \pathcond and
\latestw are also readily addressed simply by unit propagation.  In \cav,
this conflict (that $X_{rw'}$ cannot happen together with $X_{rw}$) is not
immediately resolved/learnt and the SAT solver is let to explore infeasible
paths until it learns the conflict sometime in the future.  Thus, our true
concurrency based, conflict-aware semantics naturally provides a compact,
highly effective decision procedure.

\ifx 0

\newcommand{\cc}[1]{\multicolumn{1}{c@{}}{#1}}
\newcommand{\tools}{\cc{\textsc{wk}}&\cc{Nidhugg}}

\begin{table}[t]
\centering
\scalebox{1.0125}{
\begin{tabular}{@{}l@{\,\,}l@{\,}l@{\,\,}c@{}c@{\,\,}c@{}c@{\,\,}c@{}c@{}}
\hline
    \multirow{2}{*}{Benchmark} & \multirow{2}{*}{mm} &    \multirow{2}{*}{$k$} &  \multicolumn{2}{l@{}}{\seqc} & \multicolumn{2}{l@{}}{\tso{}} & \multicolumn{2}{l@{}}{\pso}\\
    \cline{4-9}
&&    & \tools & \tools & \tools\\
    \hline
apr\_1.c & sc & 5  &   t/o  & \bf{  5.69} &   t/o   & \bf{  5.40} &   t/o &   \bf{  13.17} \\
apr\_2.c & sc & 5  &   t/o & \bf{  2.13} &   t/o & \bf{  2.31} &   t/o &  \bf{  4.94} \\
dcl\_singleton.c & sc & 7  &   49.97 &   \bf{0.13} &   122.08   & \bf{  0.09} & *55.02 & \bf{*0.11} \\
dcl\_singleton.c & pso & 7  &   51.46 & \bf{0.08} &   103.83   & \bf{  0.08} &   148.36  & \bf{  0.08} \\
dekker.c & sc & 10  &   11.32 & \bf{  0.10} & *6.17 & \bf{*0.10} & *6.32 & \bf{*0.08} \\
dekker.c & tso & 10  &   11.18 & \bf{  0.10} &   14.80 & \bf{  0.10} &   13.62 & \bf{*0.08} \\
dekker.c & pso & 10  &   11.16 & \bf{  0.10} &   14.81 & \bf{  0.10} &   13.31 & \bf{  0.11} \\
fib\_false.c & sc & no  & *6.76 & \bf{*2.02} & *4.80 & \bf{*4.60} & *9.99 & \bf{*5.49} \\
fib\_false\_join.c & sc & no  & *3.11 & \bf{*0.30} & *6.19 & \bf{*0.55} & *6.34 & \bf{*0.63} \\
fib\_true.c & sc & no  & \bf{  11.56} &   21.62 & \bf{  12.01} &   60.99 & \bf{  12.06} &   73.10 \\
fib\_true\_join.c & sc & no  &   9.24 & \bf{  1.02} &   13.62 & \bf{  2.44} &   15.83 & \bf{  2.84} \\
indexer.c & sc & 5  &   203.50 & \bf{  0.09} &   221.22 & \bf{  0.09} &   233.78 & \bf{  0.09} \\
lamport.c & sc & 8  &   28.68 & \bf{  0.08} & *20.28 & \bf{*0.08} & *24.87 & \bf{*0.08} \\
lamport.c & tso & 8  &   28.76 & \bf{  0.08} &   53.22 & \bf{  0.07} &   51.05 & \bf{*0.08} \\
lamport.c & pso & 8  &   28.71 & \bf{  0.08} &   53.27 & \bf{  0.07} &   66.27 & \bf{  0.08} \\
parker.c & sc & 10  &   38.88 & \bf{  1.20} &   186.09 & \bf{*0.09} &   375.64 & \bf{*0.09} \\
parker.c & pso & 10  &   38.93 & \bf{  1.21} &   189.47 & \bf{  1.55} &   355.09 & \bf{  2.62} \\
peterson.c & sc & no  &   0.56 & \bf{  0.07} & *0.43 & \bf{*0.07} & *0.41 & \bf{*0.07} \\
peterson.c & tso & no  &   0.56 & \bf{  0.07} &   0.59 & \bf{  0.07} & *0.40 & \bf{*0.07} \\
peterson.c & pso & no  &   0.56 & \bf{  0.07} &   0.59 & \bf{  0.07} &   0.79 & \bf{  0.07} \\
pgsql.c & sc & 8  &   227.89 & \bf{  0.08} &   421.65 & \bf{  0.07} & *13.47 & \bf{*0.10} \\
pgsql.c & pso & 8  &   227.50 & \bf{  0.08} &   581.28 & \bf{  0.07} &   t/o & \bf{  0.07} \\
pgsql\_bnd.c & sc & no  & \bf{  29.15} &   64.09 & \bf{  40.94} &   75.94 & *3.62 & \bf{*0.10} \\
pgsql\_bnd.c & pso & no  & \bf{  29.14} &   66.26 & \bf{  50.78} &   79.16 & \bf{  64.44} &   93.99 \\
stack\_safe.c & sc & no  &   256.31 & \bf{  0.30} &   395.62 & \bf{  0.31} &   578.83 & \bf{  0.38} \\
stack\_unsafe.c & sc & no  & *4.72 & \bf{*0.11} & *5.18 & \bf{*0.11} & *6.40 & \bf{*0.11} \\
szymanski.c & sc & no  &   0.76 & \bf{  0.08} & *0.70 & \bf{*0.13} & *0.66 & \bf{*0.07} \\
szymanski.c & tso & no  &   0.75 & \bf{  0.08} &   0.83 & \bf{  0.08} &   0.93 & \bf{*0.08} \\
szymanski.c & pso & no  &   0.76 & \bf{  0.08} &   0.88 & \bf{  0.08} &   0.95 & \bf{  0.08} \\
\hline
\end{tabular}
}
\caption{\tc~(\textsc{wk}) vs.~\smc}
\label{tab:eval:tc-smc}
\begin{tablenotes}
	\item [$*$] = ``time to first error''
	\item [$!$] = crash
	\item [$t/o$] = timeout
\end{tablenotes}
\end{table}

\fi

\ifx 0

We now proceed to compare our work with \smc~\cite{smc}. This work proposes
a new partial order based canonical representation, called chronological
traces to model input programs.  This trace based model then is used to
model check the input program.  \smc can only model shared memory programs
with deterministic threads; our approach does not have such restrictions. 
We were unable to run Litmus and SV-COMP~2015 benchmarks on \smc as it crashed
on almost all program instances; we are looking into it currently.  But we
were able to compare on the programs that \smc was tested on;
\cref{tab:eval:tc-smc} gives the results.  The initial results for \smc are
promising and we are currently looking into adopting chronological traces
(as opposed to unprocessed $\ppo$) to build event structures for
deterministic programs.

\begin{figure*}
\begin{subfigure}{0.33\textwidth}
\begin{tikzpicture}[domain=1e-2:1e2, scale=0.7]
\centering
\begin{loglogaxis}[%
	xlabel=FDR3 (GB),
	ylabel=SAT (GB),
	compat = 1.3,
	legend entries={SAT, UNSAT},
	legend style={at={(.27,.7)}},
	ylabel shift = -10 pt,
	xlabel shift = -3 pt,
	xmin = 1e-2, xmax=1e2,
	ymin = 1e-2, ymax=1e2,
	xtick={1e-1, 1e0, 1e1},
	ytick={1e-1, 1e0, 1e1},
	scatter/classes={%
		SATISFIABLE={mark=triangle*,red},%
		UNSATISFIABLE={mark=x,blue}%
		},
	]
	\addlegendimage{only marks, mark=triangle*,red}
	\addlegendimage{only marks, mark=x,blue}
\addplot[smooth, thick] {x};
\addplot[dashdotted,thin] {47.15};
\node at (axis cs:3e-2,47.15) [above] {47GB};
\draw [dashdotted, thin]  ({axis cs:47.15,0}|-{rel axis cs:0,0}) -- ({axis cs:47.15,0}|-{rel axis cs:0,1});
\end{loglogaxis}
\end{tikzpicture}
\captionof{figure}{Memory: \mpitocsp vs.~\definder}
\label{fig:eval:definder-mpitocsp-maxrss}
\end{subfigure}
\end{figure*}
\fi

\ifx
\begin{figure}
\centering
\scalebox{.9}{
\begin{tabular}{@{}c|c|c|c|c|c|cc@{}}
	& \multicolumn{3}{c|}{Litmus} & \multicolumn{3}{c}{SV-COMP 2015}
\\
\cline{2-7}
	&	\seqc & \tso & \pso &\seqc & \tso & \pso
\\
\hline
loc         &  63  & 63 &63 & 888 	& 801 	& 810     \\
unwind 		& -nil- & -nil- & -nil- & 6    	& 6 	& 6    \\
\hline
\multirow{2}{*}{\#vars}\hspace{0.25em}po\hspace{-0.5em}
&  3006  & 5346   & 7037  & 1.4e6 &   268359  &  298467      \\
\hspace{2.25em}tc\hspace{-0.5em} 
& 3200 & 3201 & 3226 & 1.5e6 & 187136 & 203134\\
\hline
\multirow{2}{*}{\#cons}\hspace{-0.5em}
\hspace{0.25em}po\hspace{-0.5em}
&  10105 &  20239 & 28437     & 6.9e6 & 1.2e6 & 1.4e6\\
\hspace{3.25em}tc\hspace{-0.75em}
& 10903 & 10908 & 11034      &   7.9e6 &  1.1e6 & 1.1e6
\\
\hline
\multirow{2}{*}{\#ssa}\hspace{0.25em}po\hspace{-1em}
&  322 & 481 &359 &70265 & 38656 & 39433\\
\hspace{1.75em}tc\hspace{-1em}
&266 & 266 & 268 &6729 & 5537 & 6729\\
\end{tabular}
}
\label{tab:eval:bnchmarks}
\caption{Specifics About the Benchmarks}
\end{figure}

The \cref{tab:eval:bnchmarks} presents the features of the benchmarks. All
the figures presented are averages: in case of Litmus, they average over
5804 programs and with svcomp, they average over 600 programs.  The $loc$
and $unwind$ depicts the number of lines and the BMC loop bound: the -nil-
stands for no loop unrolling.  Litmus programs are loop-free and require no
unrolling.  We use 6 as the unrolling bound in all our experiments.  The two
following rows, each with two sub-rows, depict the number of variables
(\emph{\#vars}) and number of constraints (\emph{\#cons}) coming from the
our approach and the work that is closest to us, ~\cite{cav13}; $\#ssa$
depicts the number of \ssa steps.

\fi

\section{Related Work}
\label{sec:rel}

We give a brief overview of work on program verification under weak memory
models with particular focus on assertion checking.  Finding defects in
shared memory programs is known to be a hard problem.  It is non-primitive
recursive for \tso and it is undecidable if read-write or read-read pairs
can be reordered~\cite{wmm:undecide}.  Avoiding causal loops restores
decidability but relaxing write atomicity makes the problem undecidable
again~\cite{wmm:loopfree}.

Verifiers for weak memory broadly come in two flavours: the
``operational approach'', in which buffers are modelled
concretely~\cite{wmm:fence:hw1, wmm:fence:hw2, wmm:fence:hw3, wmm:fence:hw4,
wmm:fence:hw5}, and the ``axiomatic approach'', in which the observable
effects of buffers are modelled indirectly by (axiomatically) constraining
the order of instruction executions~\cite{wmm:fence:po1, wmm:ver:po2, cav13,
smc}.  The former often relies on interleaving semantics and employs
transition systems as the underlying mathematical framework.  The later
relies on independence models and employs partial orders as the mathematical
foundation.  The axiomatic method, by abstracting away the underlying
complexity of the hardware, has been shown to enable the verification of
realistic programs.  Although we do \emph{not} use partial orders, our true
concurrency based approach falls under the axiomatic approach.

These two approaches have been used to solve two distinct, but related
problems in weak memory.  The first one is finding assertion violations that
arise due to the intricate semantics of weak memory; this is the problem we
address as well.  The other is the problem of fence insertion.  Fence
insertion presents two further sub-problems: the first is to find a
(preferably minimal, or small enough) set of fences that needs to be
inserted into a weak memory program to make it sequentially
consistent~\cite{wmm:fence:sc1, wmm:fence:po1, wmm:fence:hw2}; the second is
to find a set of fences that prevents any assertion violations caused by
weak memory artefacts~\cite{wmm:fence:hw4, wmm:fence:ver, checkfence,jk2015-fm}.

There are three works ---~\cite{cav13, smc, wmm:ver} --- that are very close
to ours.  The closest to our work, \cite{cav13}, was discussed
in~\cref{sec:eval}.  \smc~\cite{smc} is promising but can only handle
programs without data nondeterminism.  The work in \cite{wmm:ver} uses code
transformations to transform the weak memory program into an equivalent SC
program, and uses \seqc-based to tools to verify the original program.

There are further, more complex memory models. Our approach can be used
directly to model RMO.  However, we currently cannot handle POWER and ARM
without additional formalisms.  Recent work~\cite{batty15-esop} studies the
difficulty of devising an axiomatic memory model that is consistent with the
standard compiler optimizations for C11/C++11.  Such fine-grained handling
of desirable/undesirable thin-air executions is outside of the scope of our
work.

\vspace{-1ex}
\section{Conclusion}
\label{sec:conc}

We presented a bounded static analysis that exploits a conflict-aware true
concurrency semantics to efficiently find assertion violations in modern
shared memory programs written in real-world languages like~C.  We believe
that our approach offers a promising line of research: exploiting event
structure based, truly concurrent semantics to model and analyse real-world
programs.  In the future, we plan to investigate more succinct intermediate
forms like Shasha-Snir traces to cover the Java or C++11 memory model and to
study other match-related problems such as lock/unlock or malloc/free.

\paragraph{Acknowledgements}

We would like to thank the reviewers and Michael Emmi for their constructive
input that significantly improved the final draft.  Ganesh Narayanaswamy is
a Commonwealth Scholar, funded by the UK government.  This work is supported
by ERC project 280053.


\bibliographystyle{plain}
\bibliography{ref}~
%
%





\appendix
\section{Artefact description}


\subsection{Abstract}

As an artefact, we submit \tc | a bounded model checker for the safety verification of programs under various memory models. This tool implements the encoding described in Section~\ref{sec:enc} and shows better performance as compared to the state of the art, partial order based \cav tool. \tc was tested on multiple x86\_64 Linux machines running Ubuntu and Fedora. Our artefact uses MiniSAT as the underlying SAT solver. Detailed hardware and software requirements have been given in the following sections. Our publicly available artefact provides automated scripts for building, for reproducing our results, and also for generating the plots included in the paper. Apart from the noise introduced due to run-time related variation and non-determinism, we expect that the overall trends of an independent evaluation by evaluation committee to match with those shown in the paper.


\subsection{Description}

\subsubsection{Check-list (artefact meta information)}


{\small
\begin{itemize}
  \item {\bf Algorithm: A novel propositional encoding}
  \item {\bf Program: Litmus and SV-COMP15 public benchmarks, both included.}
  \item {\bf Compilation: g++ 4.6.x or higher, Flex and Bison, and GNU make version 3.81 or higher.}
  \item {\bf Transformations: None}
  \item {\bf Binary: \tc and \cav binaries are included in VM image along with the source code.}
  \item {\bf Data set: None}
  \item {\bf Run-time environment: The artefact is well tested on 64-bit x86 Linux machines running Ubuntu version 14.04 or higher, and Fedora release 20 or higher.}
  \item {\bf Hardware: 64-bit x86 Linux PC}
  \item {\bf Run-time state: Should be run with minimal other load}
  \item {\bf Execution: Sole user with minimal load from other processes}
  \item {\bf Output: We produce all the plots used in the original submission.}
  \item {\bf Experiment workflow: Readme.md}
  \item {\bf Publicly available?: Yes }
\end{itemize}
}

\subsubsection{How delivered}
The artefact and all the necessary benchmarks can be obtained by cloning the artefact repository using the following command.

{\tt git clone https://github.com/gan237/walcyrie}


\subsubsection{Hardware dependencies}
The artefact is well-tested on 64-bit x86 machines.

\subsubsection{Software dependencies}
The artefact is well-tested on 64-bit Linux machines (Ubuntu 14.04+, Fedora 20+). Other software dependencies include g++-4.6.x or higher, flex, bison, MiniSAT-2.2.0, Perl-5, libwww-perl version 6.08 (for the lwp-download executable), GNU make version 3.81 or higher, patch, gnuplot, awk, sed, epstopdf and pdf90. The Readme.md has a comprehensive list of dependencies.


\subsection{Installation}
The artefact can be downloaded using the following command:

{\tt git clone https://github.com/gan237/walcyrie.git}

The installation instructions are given at \url{https://github.com/gan237/walcyrie/blob/master/README.md}.

We also provide a ready to use virtual machine image containing the artefact at:

{\tt http://www.cprover.org/wmm/tc/ppopp16}


\subsection{Experiment workflow}
Steps to reproduce the results presented in the paper are described
in the README.md. 

\subsection{Evaluation and expected result}
We produce all the plots used in the original submission. 
The steps to produce plots are also explained in the README.md file 
supplied with the artefact. We expect that barring the noise that
may be introduced due to variation in the runtime environment, the
trends produced by the plots should show better performance
for \tc as compared to \cav as mentioned in the paper.





\end{document}